\documentclass[%
 aip,
 amsmath,amssymb,
 reprint,%
]{revtex4-1}

\usepackage{graphicx}
\usepackage{dcolumn}
\usepackage{bm}

\usepackage[utf8]{inputenc}
\usepackage[T1]{fontenc}
\usepackage{mathptmx}
\usepackage[none]{hyphenat}
\usepackage{graphicx}
\usepackage{algpseudocode}
\newcounter{algorithm}
\newenvironment{algorithm}[1][t!]{%
  \begin{figure}[#1]%
  \refstepcounter{algorithm}%
  \begin{minipage}{\columnwidth}%
  \hrule height .8pt depth 0pt\relax
  \kern 2pt
  \def\caption##1{%
    \noindent\textbf{Algorithm \thealgorithm}\ ##1\par
    \kern 2pt\hrule\kern 2pt
  }%
}{%
  \kern 2pt\hrule\relax
  \end{minipage}%
  \end{figure}%
}
\usepackage{stmaryrd}

\usepackage[colorlinks=true,
            linkcolor=blue,
            citecolor=blue,
            urlcolor=blue,
            hyperfootnotes=false]{hyperref}

\usepackage{hypcap} 

\makeatletter
\def\present@FM@footnote#1#2{%
  \begingroup
    \csname c@\@mpfn\endcsname#1\relax
    \def\@thefnmark{\frontmatter@thefootnote}%
    \par\normalfont\small
    \frontmatter@makefntext{#2}\par
  \endgroup
}%
\makeatother
\begin{document}

\title[Shortest-path percolation and structural robustness of quantum internet networks]{Shortest-path percolation and structural robustness of quantum internet networks}

\author{Victor H. F. Francheto}
\email{victor.h.f.francheto@gmail.com}
\author{Thomas Peron}%
\email{thomas.peron@usp.br}
\affiliation{Institute of Mathematics and Computer Science, University of São Paulo, São Carlos, São Paulo 13566-590, Brazil}%

\date{\today}

\begin{abstract}
We study shortest-path percolation on spatial networks that model the quantum internet based on optical fibers. In this process, the removal of paths describes the consumption of the entangled links used for communication. We compare finite and unrestricted path-length bounds for networks whose average degree either increases with system size or remains approximately constant. We show that networks become more robust as the average degree increases, whereas networks with approximately constant average degree exhibit a data collapse for the fixed finite bounds considered. Using Monte Carlo sampling of the shortest-path-length distribution, we estimate an equivalent path cost, a finite bound that approximately reproduces the degradation of the network in the unrestricted case. For networks with fixed average degree, this cost grows approximately as the square root of the number of nodes. Increasing the bound beyond this value requires additional computation time without appreciably changing the percolation dynamics. As a complementary result, for Erd\H{o}s-Rényi random networks, we find an empirical relation between the initial average degree and the approximately common crossing point of the percolation curves for different path-length bounds.
\end{abstract}

\maketitle

\begin{quotation}
Quantum communication consumes entanglement resources and can therefore change the routes available between nodes in a network. Understanding how this process affects connectivity is important for the development of a quantum internet. Here we study a model of optical-fiber networks in which each successful communication request removes the entangled links along a shortest path. We show that increasing the number of connections per node makes the network more robust, and distinguish this effect from that of increasing the number of nodes at approximately constant average degree. By randomly sampling pairs of nodes, we estimate a finite limit on path length, which we call the equivalent path cost. Within the precision of our simulations, this limit yields the same overall loss of connectivity as the process in which paths of any length can be used. When the average number of connections per node remains approximately constant, the equivalent cost grows approximately as the square root of the number of nodes. This finite limit reduces computation time without appreciably changing the observed network fragmentation. We also find an empirical relation for random networks describing the fraction of removed links at which different path-length limits yield nearly the same connectivity. Our results show how network structure and restrictions on communication paths affect the resilience of quantum networks.
\end{quotation}

\section{Introduction}
Over the past two decades, network science has become a general framework for the quantitative analysis of complex systems across statistical physics, computer science, biology, and sociology \cite{newman2018networks, vanderHofstad2017, Li_2021}. By focusing on the structure and dynamics of heterogeneous interactions, it provides tools for identifying emergent collective behavior and relating it to the organization of the underlying network. In this context, percolation theory offers a natural framework for characterizing the formation and fragmentation of large-scale connectivity \cite{stauffer1992introduction,Li_2021}. Concepts such as the giant connected component, finite-size scaling, and mean-field approximations are therefore widely used to assess structural robustness and dynamical processes in complex networks~\cite{Dorogovtsev_2008,dorogovtsev2022nature}.

Percolation theory originated from the study of macroscopic connectivity emerging from microscopic links in disordered systems~\cite{stauffer1992introduction}. Early formulations appeared in gelation theory for branched polymers, where the formation of an infinite component corresponds to a connectivity transition \cite{Flory1941, Stockmayer1944}. The theory was later formalized through lattice models of fluid flow in disordered media, including porous and labyrinth-like structures \cite{BroadbentHammersley1957}. These developments established the foundations of percolation theory, subsequently systematized in classical treatments \cite{stauffer1992introduction}.

The integration of quantum information technologies with network science has led to a class of systems known as quantum networks \cite{Kimble2008}. These infrastructures exploit quantum phenomena such as entanglement and superposition to transmit and process information, introducing challenges associated with nonlocal interactions, quantum correlations, and physical constraints on connectivity \cite{Piani_2009, Sasmal_2023}. Despite rapid experimental progress, the large-scale realization of a quantum internet remains limited by physical fragility, resource scarcity, and network-level robustness \cite{Wehner2018}. Quantum channels are sensitive to losses, decoherence, and imperfections in entanglement generation, while high-fidelity entangled links, quantum memories, and photon transmission capacity remain limited resources \cite{Sangouard2011, Pirandola_2017}. Consequently, the reliability of large quantum networks depends not only on the local performance of individual devices but also on global topology and routing dynamics. Understanding how communication and resource consumption affect network connectivity is therefore essential for assessing the scalability and resilience of future quantum communication systems.

One of the main applications of quantum networks is quantum key distribution (QKD), which uses quantum mechanics to provide information-theoretic security \cite{RevModPhys.74.145}. Other examples include high-precision clock synchronization \cite{clocks}, quantum-enhanced sensing, and the secure delegation of quantum computation \cite{Broadbent_2009, PhysRevA.96.012303}. Quantum networks also allow the study of fundamental phenomena, such as entanglement distribution over large distances, quantum teleportation, and nonlocal correlations \cite{RevModPhys.86.419}.

In network science, percolation provides a canonical description of robustness and fragmentation through the emergence or collapse of the giant component under random failures or targeted attacks~\cite{Li_2021,dorogovtsev2022nature}. In quantum networks, entanglement percolation extends this idea by asking whether local operations and classical communication can transform non-maximally entangled links into usable connections capable of sustaining large-scale entangled clusters \cite{Acin2007}. Extensions to complex networks and multipartite strategies have shown that quantum protocols can modify percolation thresholds and global connectivity relative to classical approaches \cite{Cuquet2009, Perseguers2010}.

More general percolation models have incorporated structural and functional constraints relevant to realistic networks, including $k$-core organization \cite{Dorogovtsev_2006}, interdependent systems with cascading failures \cite{Buldyrev_2010}, and multilayer or multiplex architectures \cite{Boccaletti2014, Kivela2014}. Higher-order interactions represented by hypergraphs and simplicial complexes can further alter connectivity and critical behavior \cite{Majhi2022}. Related approaches have also been applied to infrastructure systems, such as urban congestion, where edge removal according to quality or capacity reveals the formation and impact of congested clusters \cite{Li2014PercolationTI}. These extensions illustrate how constraints on connectivity and resource availability can qualitatively change network degradation.

The quantum internet is a spatially distributed communication infrastructure in which nodes are connected by quantum channels and entanglement can be established over long distances \cite{Kimble2008, Wehner2018}. From a network perspective, shortest paths play a central operational role because entanglement distribution, teleportation, and routing generally favor routes that minimize an effective cost associated with physical distance, optical losses, or fidelity degradation. Shorter paths therefore tend to provide more reliable communication. At the same time, the spatial embedding of optical-fiber networks introduces strong constraints on link lengths and reliabilities, making the availability and progressive loss of short communication routes particularly relevant to global network functionality.

The Shortest-Path Percolation (SPP) model proposed in Ref.~\cite{Kim_2024} provides a natural framework for studying this degradation mechanism. At each iteration, a source-destination pair is selected, a shortest path is identified, and the edges composing that path are removed whenever its length satisfies a prescribed cost bound. In quantum communication networks, this process can be interpreted as the consumption of entangled links used to establish end-to-end communication through entanglement swapping. Successive requests therefore modify the set of routes available for future communications, directly coupling communication activity to the evolution of network connectivity. The cost bound controls the maximum number of links that can be consumed in a single request and thus provides a simple way to investigate how restrictions on communication range affect robustness.

In this paper, we employ the SPP model of Ref.~\cite{Kim_2024} to investigate how structure, spatial organization, and the maximum allowed shortest-path cost determine the degradation of quantum internet networks. We consider optical-fiber networks in which photon losses increase with propagation distance and compare processes with limited and unrestricted shortest-path lengths. We study two spatial constructions: a fixed domain, where the density of nodes and the average degree increase with the network size, and a rescaled domain, where the average degree remains approximately constant. This comparison allows us to examine the effect of increasing connectivity on the degradation dynamics as the number of nodes grows. We also consider different strategies for selecting source-destination pairs, comparing path removals between randomly selected pairs with removals that affect targeted nodes.

The manuscript is organized as follows. In Section~\ref{sec:topology}, we introduce the spatial network model used to represent quantum internet architectures and discuss its main structural properties. In Section~\ref{sec:spp}, we present the SPP model and analyze the resulting connectivity degradation, phase transitions, pair-selection strategies, equivalent finite cost, and the effects of fixed- and rescaled-domain constructions. In Section~\ref{sec}, we analyze the common crossing point observed in the percolation curves of random networks and its dependence on network connectivity. Finally, Section~\ref{sec:conclusion} summarizes the main findings and discusses their implications for the robustness of quantum communication networks.

\section{Structure and dynamics of quantum internet networks}
\label{sec:topology}

In this section, we introduce the optical-fiber quantum network model of Ref.~\cite{Brito_2020} and describe the structural properties relevant to the SPP dynamics studied below. The original study identified a critical node density separating a fragmented regime from one containing a giant connected component and estimated the corresponding critical exponents. These results suggest that relatively few nodes can sustain large-scale connectivity in realistically sized photonic networks. Beyond the structural connectivity familiar from studies of the classical internet \cite{Albert_1999}, quantum networks can also acquire effective long-range connections through nonlocal entanglement generation and swapping. These protocols can give rise to small-world behavior even when it is absent from the underlying physical infrastructure \cite{Cuquet2009}. The resulting short typical paths are operationally relevant because they suggest that only a few entanglement-swapping operations may be required to establish entanglement between distant nodes \cite{Abane2025}. We therefore begin by distinguishing the physical fiber-optic infrastructure from the photonic connections established over it, which form the networks used in our subsequent analysis.

\subsection{Modeling the Fiber-Optics and Photonic Network}
\label{sec:fiber_photonic_nets}


We begin by uniformly distributing $N$ nodes within a disk of radius $R>0$. Following the approach in Ref.~\cite{1217279}, we simulate the optical fiber connections between these nodes using the Waxman model~\cite{12889,barthelemy2011spatial}, where each pair of nodes is connected by a fiber with probability $\Pi_{ij} = \beta e^{-d_{ij}/\alpha L}$, in which $d_{ij}$ is the Euclidean distance between nodes $i$ and $j$, $L$ is the maximum distance between any two nodes, $\alpha>0$
sets the characteristic decay length $\alpha L$, and $0 < \beta \leq 1$ governs the density of links of the network. The constants $\alpha$ and $\beta$ have been estimated in real optical networks. For example, in the US fiber-optics network, $\alpha L = 226$ km and $\beta = 1$ \cite{Durairajan}. These values are kept constant in the simulations presented here. Note that, as $d_{ij} \to \infty$, the probability $\Pi_{ij} \to 0$, meaning that the probability of a connection decreases as the distance between nodes increases.

 
Once the fiber-optics network is established, we simulate the transmission of photons through it. Photon losses increase exponentially with fiber length \cite{Gisin_2015}. Specifically, the transmissivity $p_{ij}$, defined as the fraction of energy received at the output of a fiber link between nodes $i$ and $j$, is given by \cite{Brito_2020}
\begin{equation}
p_{ij} = 10^{-\gamma d_{ij}/10},
\label{probability_1}
\end{equation}
where $d_{ij}$ (in km) is the distance between nodes $i$ and $j$, and $\gamma$ is the fiber attenuation in $\mathrm{dB/km}$. In optical fibers, losses are minimized at a wavelength of $1550\,\mathrm{nm}$, resulting in $\gamma \approx 0.2\,\mathrm{dB/km}$, which is the value adopted in our simulations. As discussed in \cite{TSUJIKAWA2005319}, even with technological advances, the fundamental loss limit for optical fibers lies between $0.095$ and $0.13\,\mathrm{dB/km}$.

If $n_p$ photons are independently sent through the fiber link $(i,j)$, the probability that at least one photon reaches the destination is
\begin{equation}
\mathcal{P}_{ij} = 1 - (1 - p_{ij})^{n_p}.
\label{probability_2}
\end{equation}

The physical fiber infrastructure is described by an adjacency matrix $A=[A_{ij}]$, where $A_{ij}=1$ indicates the existence of a physical fiber link between nodes $i$ and $j$, and $A_{ij}=0$ otherwise. The effective photonic connectivity is modeled by a random adjacency matrix $A^{(ph)}=[A^{(ph)}_{ij}]$, whose entries are Bernoulli random variables with parameter $A_{ij}\mathcal{P}_{ij}$, that is:
\begin{align}
\Pr\!\left(A_{ij}^{(\mathrm{ph})}=1 \mid A_{ij},d_{ij}\right)
= A_{ij}\mathcal{P}_{ij}.
\label{eq:photon_transmission}
\end{align}

\begin{figure}[!t]
  \centering
\includegraphics[width=\linewidth]{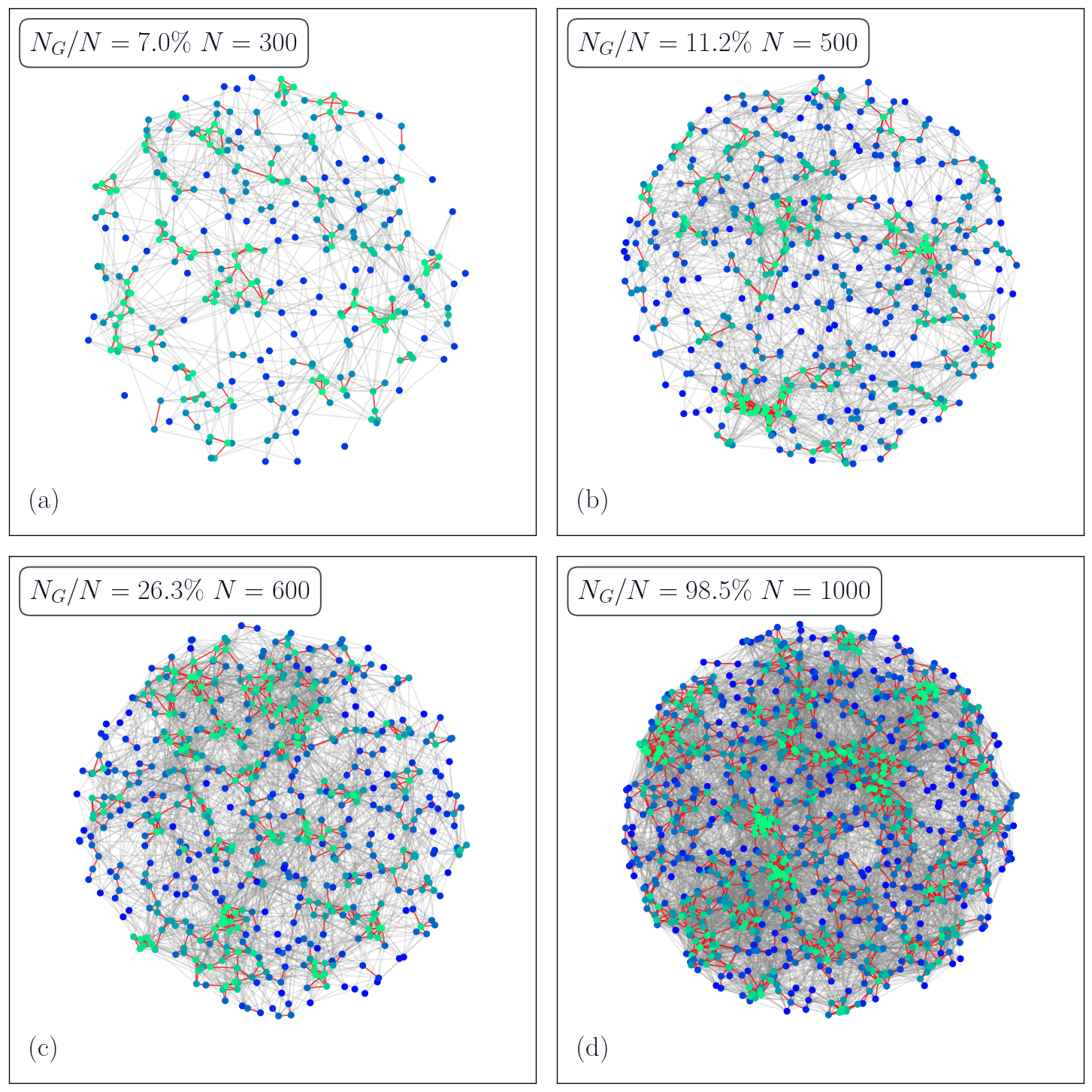}
  \caption{Examples of networks generated with the quantum internet model described in Section~\ref{sec:fiber_photonic_nets}, considering nodes distributed over a disk with fixed radius $R=1800~\mathrm{km}$. Panels (a--d) correspond to $N=300$, $500$, $600$, and $1000$ nodes, respectively. Gray edges represent the underlying fiber-optic infrastructure generated by the Waxman model, while red edges indicate the photonic links established through photon transmission [Eq.~\eqref{eq:photon_transmission}]. Node colors indicate degree, increasing from blue to green. The ratio $N_G/N$ denotes the fraction of nodes belonging to the largest connected component of the photonic network. These realizations illustrate how increasing node density within a fixed spatial domain promotes global photonic connectivity, with the largest component containing $98.5\%$ of the nodes in panel (d).}
\label{fig:quantum_network}
\end{figure}

As in Ref.~\cite{Brito_2020}, we adopt $n_p=10^3$ throughout the simulations. This choice is physically relevant because, although the transmission probability of a single photon decays exponentially with distance, the transmission of multiple photons still allows for a non-negligible probability of activating photonic links over experimentally relevant distances. Figure~\ref{fig:quantum_network} shows illustrations of networks created with the network model described above. For networks with $N=10^{3}$, approximately $98.5\%$ of the nodes belong to a giant connected component.

The connectivity observed in Fig.~\ref{fig:quantum_network} arises from photonic links whose activation depends on both the distance between nodes and the number of photons sent. To quantify the spatial range of these connections, we introduce two characteristic distance scales. The first is associated with the attenuation of a single photon in the optical fiber. Given a threshold $\varepsilon\in(0,1)$, the distance beyond which the single-photon transmission probability satisfies $p_{ij}<\varepsilon$ is obtained from $10^{-\gamma d_{ij}/10}<\varepsilon$, namely
\begin{align}
d_{ij}>d_{\mathrm{fiber}}(\varepsilon):=-\frac{10}{\gamma}\log_{10}(\varepsilon).
\end{align}
The second scale is associated with the activation of a photonic link when $n_p$ photons are sent. Defining the effective photonic range as the distance beyond which $\mathcal P_{ij}<\varepsilon$, we obtain from Eq.~\eqref{probability_2}
\begin{align}
p_{ij}<1-(1-\varepsilon)^{1/n_p}.
\end{align}
Substituting Eq.~\eqref{probability_1} into the equation above, we obtain
\begin{align}
d_{ij}>d_{\mathrm{photonic}}(\varepsilon):=
-\frac{10}{\gamma}
\log_{10}\left(1-(1-\varepsilon)^{1/n_p}\right).
\end{align}
For the parameters adopted in this work, $\gamma=0.2\,\mathrm{dB/km}$, $n_p=1000$, and the reference threshold $\varepsilon=10^{-2}$, we obtain $
d_{\mathrm{fiber}}\approx100\,\mathrm{km}$ and $d_{\mathrm{photonic}}\approx250\,\mathrm{km}$.

These values provide an interpretation of the spatial scales present in the model. While $d_{\mathrm{fiber}}$ characterizes the attenuation length of a single photon in the optical fiber, $d_{\mathrm{photonic}}$ quantifies the effective range over which multi-photon transmission still yields a non-negligible probability of activating a photonic link. Therefore, node pairs separated by distances much larger than these scales contribute negligibly to the probabilities $p_{ij}$ and $\mathcal P_{ij}$, making long-range photonic connections progressively less likely.

The existence of an effective connection range implies that photonic 
connectivity is mainly determined by the number of nodes lying within 
the interaction region of each node. The empirical average degree of a network with $N$ nodes is defined by
\begin{equation}
    \langle k\rangle =
    \frac{1}{N}\sum_{i=1}^{N} k_i,
\end{equation}
where $k_i = \sum_j A_{ij}^{(ph)}$ denotes the number of connections of node $i$ in the photonic network. Throughout 
this work, $\langle k\rangle$ refers specifically to the average degree 
of the final photonic network, that is, the network obtained after 
constructing the underlying fiber-optic infrastructure and subsequently 
establishing successful photonic connections over its fiber links. In 
particular, when the number of nodes increases while the spatial domain 
remains fixed, the node density also increases, so that more nodes fall 
within the effective photonic range. Consequently, the expected number 
of photonic neighbors and, hence, the average degree $\langle k\rangle$ 
increase. The impact of this mechanism on the network structure, 
connectivity, and formation of the giant component will be explored in 
the following sections.

\section{Shortest-Path Percolation in Quantum Internet Networks}
\label{sec:spp}

The Shortest-Path Percolation (SPP) dynamics~\cite{Kim_2024} is defined as follows. Let $t \in \mathbb{N}$ denote the iteration index of the degradation process. At each iteration $t$, a new agent generates an origin--destination request $o_t \to d_t$. The infrastructure available at that iteration is represented by an undirected and unweighted network $\mathcal{G}_t=(\mathcal{V},\mathcal{E}_t)$, composed of a fixed set of $N=|\mathcal{V}|$ nodes and an evolving set of edges $\mathcal{E}_t$. Here, $|\mathcal{E}_t|$ denotes the number of links remaining in the network after the previous path-removal operations.

If at least one path connecting $o_t$ and $d_t$ exists in $\mathcal{G}_t$, we define $Q_t$ as the length of the shortest path between the two nodes. The demand of agent $t$ is considered successfully satisfied if and only if the destination node $d_t$ is reachable from the origin node $o_t$ and the corresponding shortest-path length satisfies $Q_t \leq C$, where $C > 0$ is a tunable parameter of the model that represents a maximum admissible path length. We adopt shortest-path routing on the unweighted photonic network, measuring path lengths by the number of edges. Physical distances affect the probabilities of establishing photonic links, while route selection is based on hop count. Edge removal represents the consumption of available entangled links, which are not replenished during the simulated dynamics. The underlying fiber-optic infrastructure remains intact.

The parameter $C>0$ defines a global cost scale that regulates the accessibility of shortest paths in the network. By fixing $C$, we impose an effective constraint on which routes remain functionally available, such that the percolation dynamics reflects not only structural connectivity but also limitations associated with the accumulated cost along the paths. In this sense, $C$ can be interpreted as an \emph{effective connectivity budget}, above which shortest paths become unviable. Throughout this work, $C=N$ ($C \rightarrow\infty$) denotes the unrestricted case.

\begin{figure}[!htbp]
  \centering
  \includegraphics[scale=0.48]{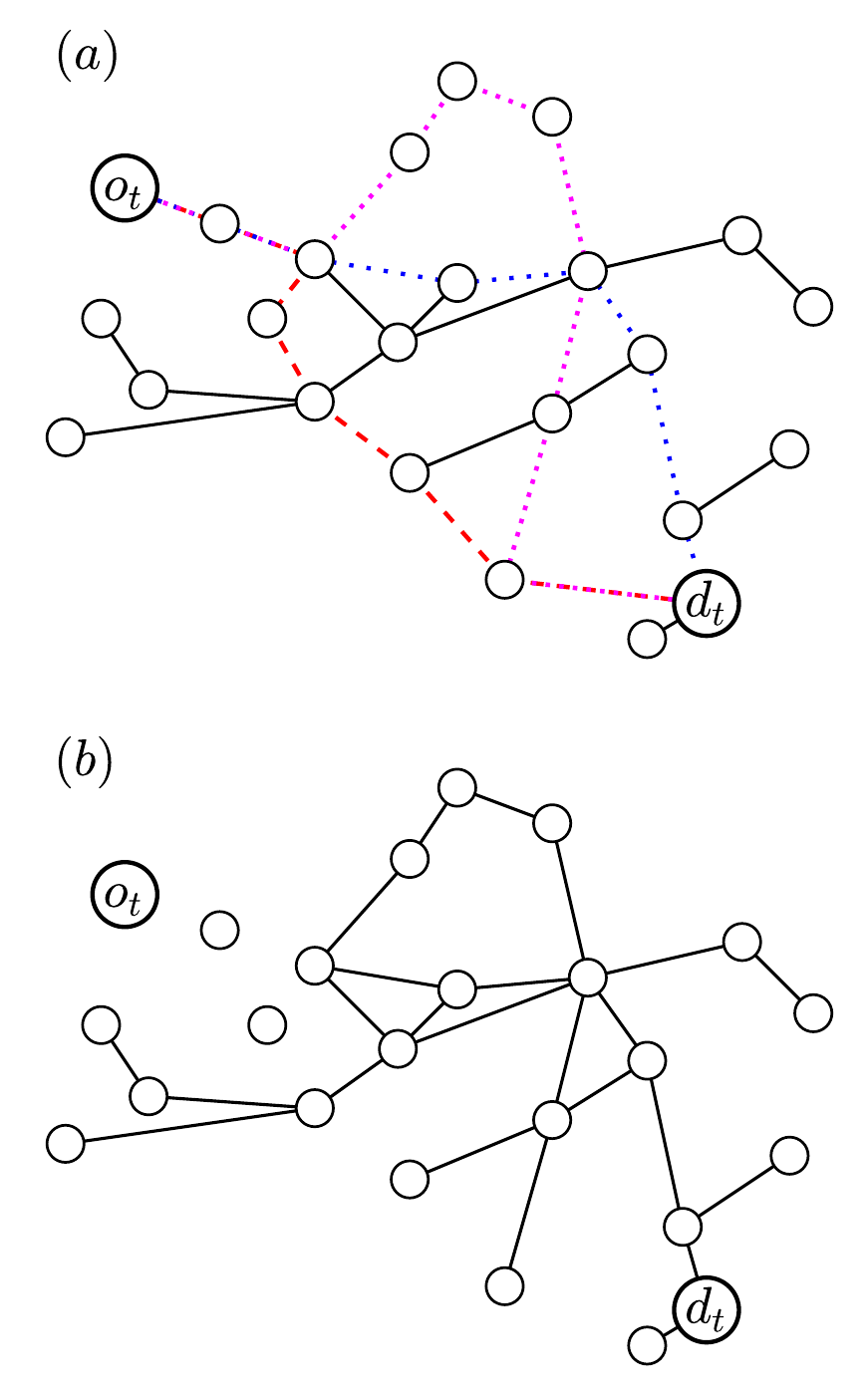}
  \caption{\textbf{Shortest-Path Percolation (SPP) model:}
  $(a)$ At time step $t$, an agent requests an origin--destination pair
  $o_t \to d_t$. In the current network $\mathcal{G}_t$, there exist two
  distinct shortest paths connecting $o_t$ and $d_t$, both with length
  $Q_t=7$, highlighted in blue and red. In addition, a longer alternative
  path with length $Q_t'=9$ is shown in magenta, representing a suboptimal
  route.
  $(b)$ If the admissible threshold satisfies $C\geq 7$, the agent's demand
  is fulfilled by randomly selecting one of the shortest paths, whose edges
  are removed from the network to model resource consumption. In the
  example, the red path is selected, and its removal fragments the network
  into multiple disconnected components, illustrating how successive
  shortest-path removals degrade global connectivity, even when longer
  alternative routes remain available.}
  \label{fig:spp_diagram}
\end{figure}

If a shortest path satisfying this condition is identified, all edges along that path are removed from $\mathcal{G}_t$. If multiple shortest paths exist, one is selected randomly. We denote the selected path by $(o_t=i_1,i_2,\ldots,i_{Q_t+1}=d_t)$. Formally, $ \mathcal{E}_t \setminus \bigcup_{q=1}^{Q_t}(i_q, i_{q+1}) \mapsto \mathcal{E}_{t+1}$. See Fig.~\ref{fig:spp_diagram} for an illustration of this process.

If there is no path between $o_t$ and $d_t$ in $\mathcal{G}_t$, or if the corresponding shortest-path length satisfies $Q_t > C$, no edge is removed. In both cases, the network remains unchanged, $\mathcal{G}_{t+1} \mapsfrom \mathcal{G}_t$, and the time index is advanced to the next step. This procedure is repeated until new demands can be realized.

The control parameter of the SPP model is $t>0$, which represents the number of source--destination pairs that have been requested in the network. A second control parameter is the fraction of edges removed, denoted by $p=p(t)$. This parameter enables direct comparisons between the SPP model and the classical percolation models. That is, the fraction of removed edges after $t$ iterations is defined as
\begin{equation}\label{paramter_p}
p(t) = 1 - \frac{E_{t+1}}{E_1},
\end{equation}
where $E_{t+1}=|\mathcal{E}_{t+1}|$ denotes the number of edges in the graph $\mathcal{G}_{t+1}$ after $t$ shortest-path requests.

The behavior of the SPP model depends on the structure of the initial network $\mathcal{G}_1$ and the agent’s demand process. In our case, we assume that the network $\mathcal{G}_1$ is an instance of a quantum internet network. Furthermore, we assume that the origin--destination pair $(o_t, d_t)$ demanded by agent $t$ is selected uniformly at random in one case, and, in another case, the origin node is chosen as the highest-degree node while the destination node is selected at random.

\begin{algorithm}[t!]
\caption{Shortest-Path Search}
\label{alg:spp1}
\begin{algorithmic}[1]

\State \textbf{Input}: Graph $\mathcal{G}$; origin $o$; destination $d$; budget $C$.
\State \textbf{Output}: Shortest-path data $P_i$ and $s_i$.

\State Run BFS from $o$ up to distance $C$, while building $P_i$ and $s_i$.

\State For each reached node $i$, let $P_i$ be the set of nodes connected to $i$
that are one step closer to $o$, and let $s_i$ be the number of shortest paths
from $o$ to $i$.

\If{$d$ is not reached}
    \State \Return \textbf{No feasible path}.
\EndIf

\Return $P_i$ and $s_i$.

\end{algorithmic}
\end{algorithm}

Following the methodology of Ref.~\cite{Kim_2024}, we perform a systematic numerical analysis based on a large number of independent simulations for each combination of the parameters $N$ and $C=1, 2 , 3$, and $N$. For each realization, a quantum network is generated and the SPP model is applied to evaluate the evolution of the network connectivity.

\begin{algorithm}[t!]
\caption{Random Shortest-Path Sampling}
\label{alg:spp2}
\begin{algorithmic}[1]

\State \textbf{Input}: Origin $o$; destination $d$; sets $P_i$; values $s_i$.
\State \textbf{Output}: Path $\pi$.

\State $\pi \leftarrow [d]$ and $i \leftarrow d$.

\While{$i \neq o$}
    \State Select $j \in P_i$ with probability proportional to $s_j$.
    \State Append $j$ to $\pi$ and set $i \leftarrow j$.
\EndWhile

\State Reverse $\pi$.

\Return $\pi$.

\end{algorithmic}
\end{algorithm}

The results obtained for quantum networks are compared with those for Erd\H{o}s--R\'enyi random graphs (ER) of the same size and average degree. Since ER graphs provide a standard null model for percolation studies \cite{Newman_2001, Maslov_2002, Milo2002}, this comparison allows us to identify structural effects associated with the spatial organization and physical constraints of quantum networks.

\subsection{Simulating the Shortest-Path percolation model}

For the sake of clarity, we present the complete SPP dynamics~\cite{Kim_2024} in Algorithms~\ref{alg:spp1}--\ref{alg:spp3}. The shortest-path search and random path sampling are described separately in Algorithms~\ref{alg:spp1} and~\ref{alg:spp2}, while the path removal algorithm is shown in Algorithm~\ref{alg:spp3}. The implementation is available at~\cite{francheto2026spp}.

At each iteration $t$, an origin--destination pair $(o_t,d_t)$ is selected according to the prescribed pair-selection rule. Algorithm~\ref{alg:spp1} searches the current graph $\mathcal{G}_t$ up to depth $C$. If no path of length at most $C$ exists between the selected nodes, the network remains unchanged, $\mathcal{G}_{t+1}=\mathcal{G}_t$. Otherwise, Algorithm~\ref{alg:spp2} samples one of their shortest paths uniformly at random, and all edges along this path are removed to obtain $\mathcal{G}_{t+1}$. The iteration counter advances after every request, regardless of whether a path is removed.

The search in Algorithm~\ref{alg:spp1} uses breadth-first search (BFS) from the origin $o$. The shortest-path alternatives are represented by a directed acyclic graph (DAG), obtained by retaining edges between consecutive BFS layers and orienting them away from $o$. For each reached node $i$, the predecessor set $P_i$ and the number of shortest paths $s_i$ from $o$ to $i$ are recorded. If the destination $d$ is reached, Algorithm~\ref{alg:spp2} constructs a path by walking backward from $d$ to $o$, selecting each predecessor $j\in P_i$ with probability $s_j/s_i$. Reversing the resulting node sequence yields a uniformly sampled shortest path from $o$ to $d$.

\begin{algorithm}[t!]
\caption{Shortest-Path Percolation with Budget Limit}
\label{alg:spp3}
\begin{algorithmic}[1]

\State \textbf{Input}: Initial graph $\mathcal{G}_1=(\mathcal{V},\mathcal{E}_1)$;
budget $C>0$.
\State \textbf{Output}: Final graph $\mathcal{G}$.

\State $t \leftarrow 1$.

\While{the stopping condition is not reached}

    \State Select an origin--destination pair $(o_t,d_t)$.

    \State Apply Algorithm~\ref{alg:spp1} to $(o_t,d_t)$.

    \If{$d_t$ is reached within distance $C$}

        \State Apply Algorithm~\ref{alg:spp2} to obtain a path $\pi_t$.

        \State Remove all edges of $\pi_t$ from $\mathcal{E}_t$.

        \State Set $\mathcal{G}_{t+1}$ according to the updated edge set.

    \Else

        \State $\mathcal{G}_{t+1}\leftarrow\mathcal{G}_t$.

    \EndIf

    \State $t\leftarrow t+1$.

\EndWhile

\Return $\mathcal{G}$.

\end{algorithmic}
\end{algorithm}

\subsection{Giant component in the shortest-path percolation dynamics}

To investigate the macroscopic connectivity of quantum networks under the SPP dynamics, we denote by $N_G$ the number of nodes in the largest connected component at each stage of the process. The ratio $N_G/N$ therefore measures the fraction of nodes belonging to this component, where $N$ is the fixed total number of nodes. We track this quantity as a function of the removed-edge fraction $p$, defined in~\eqref{paramter_p}, to characterize connectivity degradation.

In quantum networks embedded within a spatially bounded region, the average degree depends on the node density. In our model, the network is embedded in a disk of fixed radius $R>0$. Therefore, increasing the number of nodes raises the node density and, consequently, the average degree of the network \cite{Brito_2020}. As a result, even networks with a relatively small number of nodes can exhibit high connectivity because $R$ is kept constant. This regime provides a suitable framework for investigating how spatial constraints influence global connectivity, the emergence of the giant component, and the robustness of the network under link degradation.

\begin{figure*}[!tp]
  \centering

  \includegraphics[
    width=\textwidth
  ]{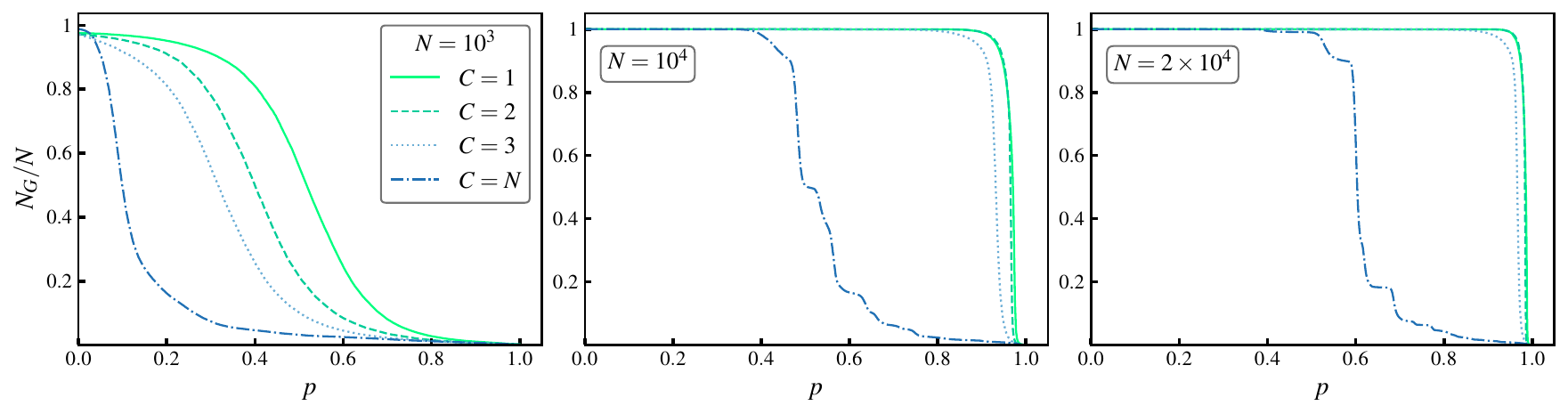}

  \caption{Shortest-path percolation in quantum internet networks embedded in a disk of fixed radius, with origin--destination pairs selected uniformly at random. The relative size of the giant component, $N_G/N$, is shown as a function of the fraction of removed edges $p$, defined in~\eqref{paramter_p}. From left to right, the panels correspond to $N=10^{3}$, $10^{4}$, and $2\times10^{4}$ nodes. The curves correspond to maximum allowed shortest-path lengths $C=1$, $2$, $3$, and $N$, with $C=N$ denoting the unrestricted dynamics. Each curve represents an average over $10^{3}$ independent network realizations.}
  \label{fig:spp-quantum-cluster-random}

\end{figure*}

The results shown in Fig.~\ref{fig:spp-quantum-cluster-random} correspond to the average behavior computed over these realizations. As we see in the figure, as the cost parameter $C$ increases, the degradation of the giant component becomes progressively more abrupt. This effect is weak for small networks but becomes pronounced for $N \geq 10^{3}$, where larger values of $C$ lead to a significantly faster decay of $N_G/N$.

\begin{figure*}[!tp]
  \centering
  \includegraphics[width=\textwidth]{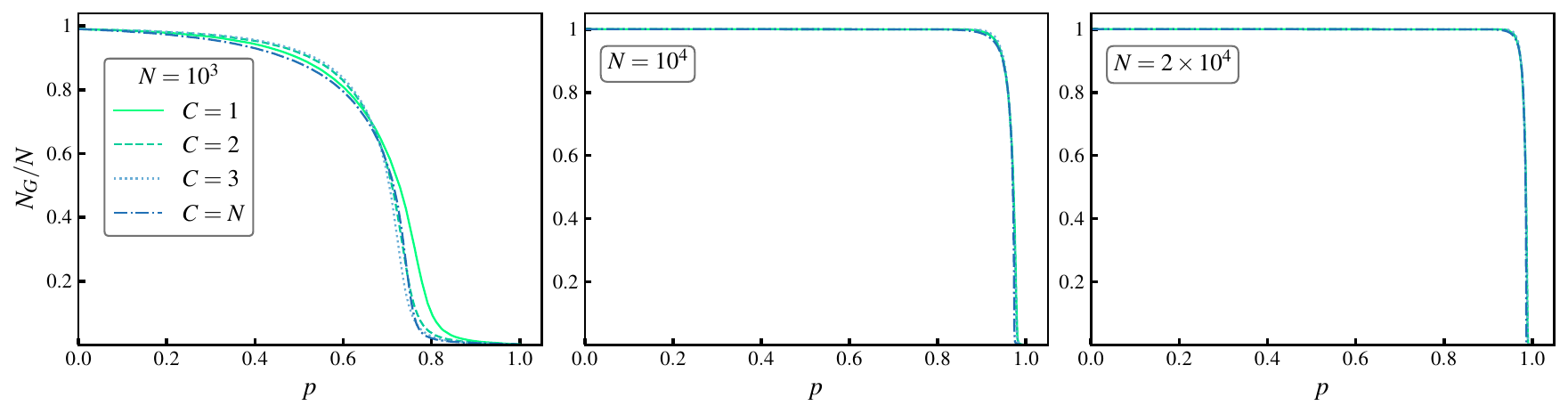}
  \caption{Shortest-path percolation in Erd\H{o}s--Rényi (ER) networks, with origin--destination pairs selected uniformly at random. The relative size of the giant component, $N_G/N$, is shown as a function of the fraction of removed edges $p$, defined in~\eqref{paramter_p}. From left to right, the panels correspond to $N=10^{3}$, $10^{4}$, and $2\times10^{4}$ nodes, with initial average degrees matched to those of the corresponding quantum networks in Fig.~\ref{fig:spp-quantum-cluster-random}. The curves correspond to maximum allowed shortest-path lengths $C=1$, $2$, $3$, and $N$, with $C=N$ denoting the unrestricted dynamics. Each curve represents an average over $10^{3}$ independent network realizations.}
  \label{fig:spp-quantum-cluster-random-er}
\end{figure*}

We next analyze the emergence of the giant component in ER random networks constructed with the same number of nodes and average degree as the quantum networks considered earlier. By matching $N$ and the initial average degree $\langle k\rangle$, we compare the robustness of spatially constrained photonic networks with that of an ER reference ensemble, in which connection probabilities are independent of spatial distance. The results are presented in Fig.~\ref{fig:spp-quantum-cluster-random-er}. As we see in the figure, as $N$ increases, the transition becomes sharper and the influence of the cost parameter $C$ becomes negligible. 

We also analyze the evolution of the giant component in the same quantum networks using a modified SPP selection rule, in which the first node is the highest-degree node and the second is selected at random. This allows us to assess how selecting a highly connected node influences the robustness of the giant component compared with the fully random selection protocol. The resulting percolation patterns remain essentially unchanged from those observed in Fig.~\ref{fig:spp-quantum-cluster-random}, with only negligible differences on average, indicating that the choice between these selection rules has little effect on the degradation dynamics. The same behavior is observed for ER networks. We hypothesize, however, that degree-based selection rules may have a more pronounced effect on SPP dynamics in networks with strongly heterogeneous degree distributions.

In the SPP dynamics with random pair selection, the quantum network defined over a fixed geometry exhibits a strong dependence on the cost parameter $C$, as illustrated in Fig.~\ref{fig:spp-quantum-cluster-random}. The main result is that the spatial constraints of the network cause the SPP dynamics to depend strongly on the cost parameter $C$. Although increasing $N$ raises the network density and the number of available paths, the probability of establishing quantum links continues to decrease with distance, causing the connections to remain predominantly local. As a consequence, communication between distant pairs of nodes relies on shortest paths composed of many short links, making entanglement distribution particularly sensitive to the limit imposed by $C$.

The $N_G/N$ curves for ER networks exhibit a much weaker dependence on the cost $C$. For each value of $N$, the different curves almost collapse onto a single one, and the position of the percolation threshold is determined essentially only by $\langle k \rangle$. As $N$ increases, the transition becomes more abrupt, but without the shift observed in quantum networks as $C$ grows.

Therefore, quantum networks are significantly less robust than ER networks under the SPP dynamics. In quantum networks, the collapse of the giant component occurs earlier and is strongly modulated by $C$. As a consequence, the network rapidly loses its ability to distribute entanglement between distant pairs of nodes as links belonging to shortest paths are removed. In contrast, random networks exhibit greater robustness and are practically independent of the cost parameter, since the abundance of alternative paths preserves global connectivity.

\subsection{Phase Transition in Quantum Internet Networks with Variable Average Degree}

To complement the analysis of the giant component of the previous section, we now examine how nodes are distributed among the remaining connected components as shortest paths are removed. The mean cluster size provides an additional measure of fragmentation and helps characterize the transition region for different values of $N$ and $C$~\cite{Kim_2024}. We focus here on the fixed-domain model, in which increasing the network size also increases the average degree.

Let the connected components at a given stage have sizes
$s_1 \le s_2 \le \cdots \le s_J$ and
$\sum_{\mu=1}^J s_\mu = N$, where $s_J$ is the size of the largest component.
The percolation strength is defined as
$P = s_J/N$, where $s_J = N_G$ denotes the size of the largest connected component.
Thus, $P = N_G/N$ measures the fraction of nodes contained in the giant component.

The mean cluster size, excluding the largest component, is defined as~\cite{stauffer1992introduction}
\begin{equation}
\label{tamanho_cluster_max}
S \,=\, \frac{\sum_{\mu=1}^{J-1} s_\mu^2}{\sum_{\mu=1}^{J-1} s_\mu}.
\end{equation}

This quantity is the expected size of the connected component containing
a node selected uniformly at random from outside the largest
component~\cite{stauffer1992introduction}. Indeed, the probability that
the selected node belongs to component $\mu$ is $s_\mu/(N-s_J)$.
Thus, $S$ is a node-weighted average of component sizes. Equivalently,
it is the ratio of the second to the first moment of the component-size
distribution obtained by sampling uniformly among the $J-1$ components
other than the largest. This weighting gives greater weight to larger components. By comparison,
the arithmetic mean
$\bar{s}=(J-1)^{-1}\sum_{\mu=1}^{J-1}s_\mu$
assigns equal weight to each component and gives the expected size of
a uniformly selected component. The two quantities therefore correspond
to different sampling procedures, that is, selecting a component uniformly or
selecting a node uniformly from outside the largest component.

\begin{figure*}[!tp]
  \centering
\includegraphics[width=\textwidth]{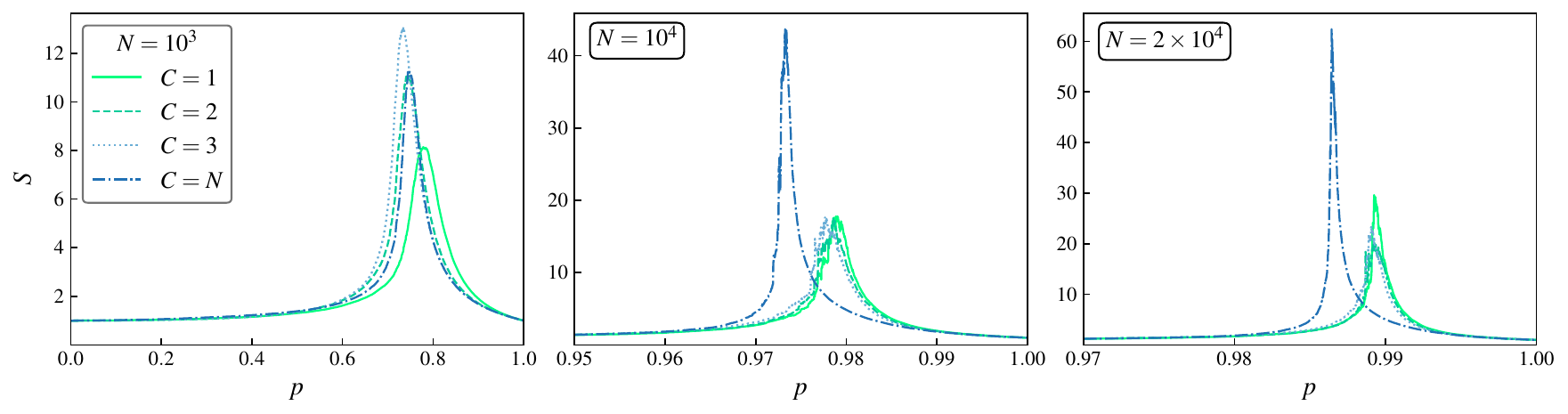}
 \caption{Average cluster size $S$ for ER networks under the SPP model. Each panel corresponds to a network size $N$. The curves correspond to maximum allowed shortest-path lengths $C=1$, $2$, $3$, and $N$, with $C=N$ denoting the unrestricted dynamics. Each curve represents an average over $10^{3}$ independent network realizations. As $N$ increases, the peaks become higher and narrower and shift toward $p=1$, indicating that fragmentation occurs after a progressively larger fraction of edges has been removed.}
 \label{fig:avarege_cluester_er}
\end{figure*}

\begin{figure*}[!tp]
  \centering
  \includegraphics[width=\textwidth]{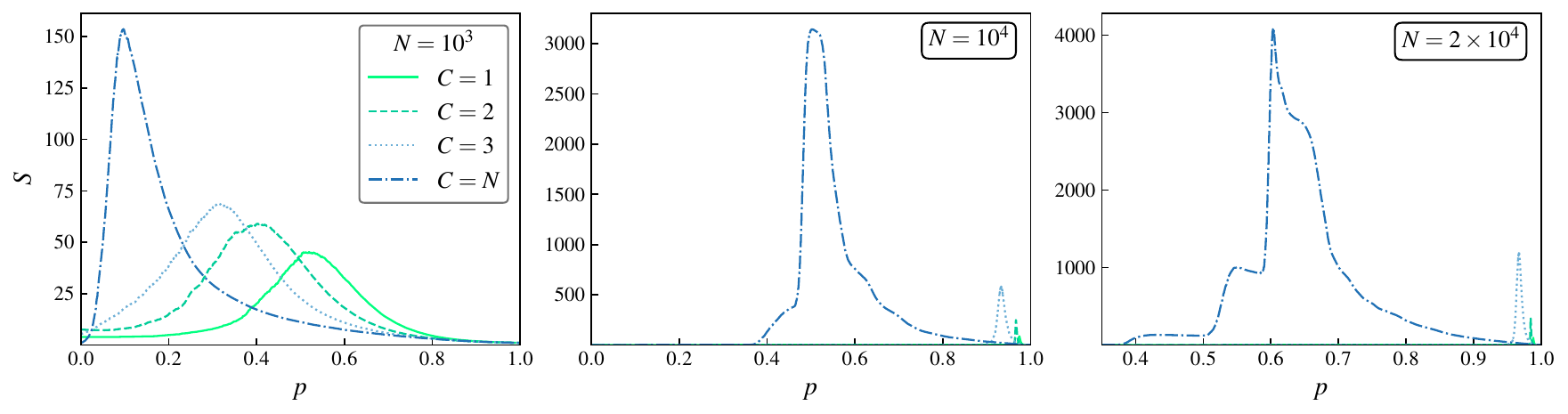}
  \caption{Average cluster size $S$ for quantum networks under the SPP model. Each panel corresponds to a network size $N$. The curves correspond to maximum allowed shortest-path lengths $C=1$, $2$, $3$, and $N$, with $C=N$ denoting the unrestricted dynamics. Each curve represents an average over $10^{3}$ independent network realizations. The peaks reflect the formation of large finite clusters as the giant component fragments.}
  \label{fig:avarege_cluester_s}
\end{figure*}

As the control parameter approaches the critical percolation threshold $p_c$, increasingly larger finite components emerge~\cite{stauffer1992introduction,sahimi1994applications}. In this regime, although a giant component has not yet formed, the presence of large finite clusters increases significantly. Since the term $\sum_{\mu=1}^{J-1} s_\mu^2$ assigns a greater weight to larger components, its contribution grows rapidly in this region, leading to an increase in $S$. In the thermodynamic limit, this quantity diverges, characterizing the critical behavior of the system.

This property establishes a direct analogy between $S$ and the susceptibility of critical systems. Whereas $P$ indicates the presence of a giant component, $S$ quantifies the fluctuations associated with finite components and captures the structural reorganization of the network in the vicinity of the percolation transition \cite{Dorogovtsev_2008}. In particular, the growth of $S$ reflects the increasing probability that a randomly chosen vertex belongs to large finite components as the system approaches the critical threshold. Together, the quantities $P$ and $S$ allow us to identify how and when the quantum network loses global connectivity under the SPP dynamics, and how this critical behavior depends on the cost parameter $C$ that determines the penalization of longer paths.

The behavior of the average cluster size $S$ for different network sizes under the quantum SPP model is shown in Fig.~\ref{fig:avarege_cluester_s}. Distinct behaviors are observed depending on the maximum path length $C$. For the unrestricted dynamics $C=N$, $S$ exhibits a maximum whose amplitude increases rapidly with the system size, while its position shifts toward smaller values of $p$. This behavior reflects the formation of large finite clusters as the giant component breaks apart. In contrast, for the constrained dynamics $C=1, 2$, and $3$, the maxima remain located near $p\approx1$, indicating that fragmentation occurs only after most node pairs have already become disconnected. As the network size increases, these maxima become progressively higher and more localized, revealing stronger finite-size effects and a more abrupt fragmentation process. ER networks exhibit a percolation transition with weak dependence on the cost parameter $C$, as shown in Fig.~\ref{fig:avarege_cluester_er}. In quantum networks, spatial embedding and the path-length constraint produce a transition region whose position and width depend on $C$.

\begin{figure}[!htbp]
  \centering
  \includegraphics[width=\linewidth]{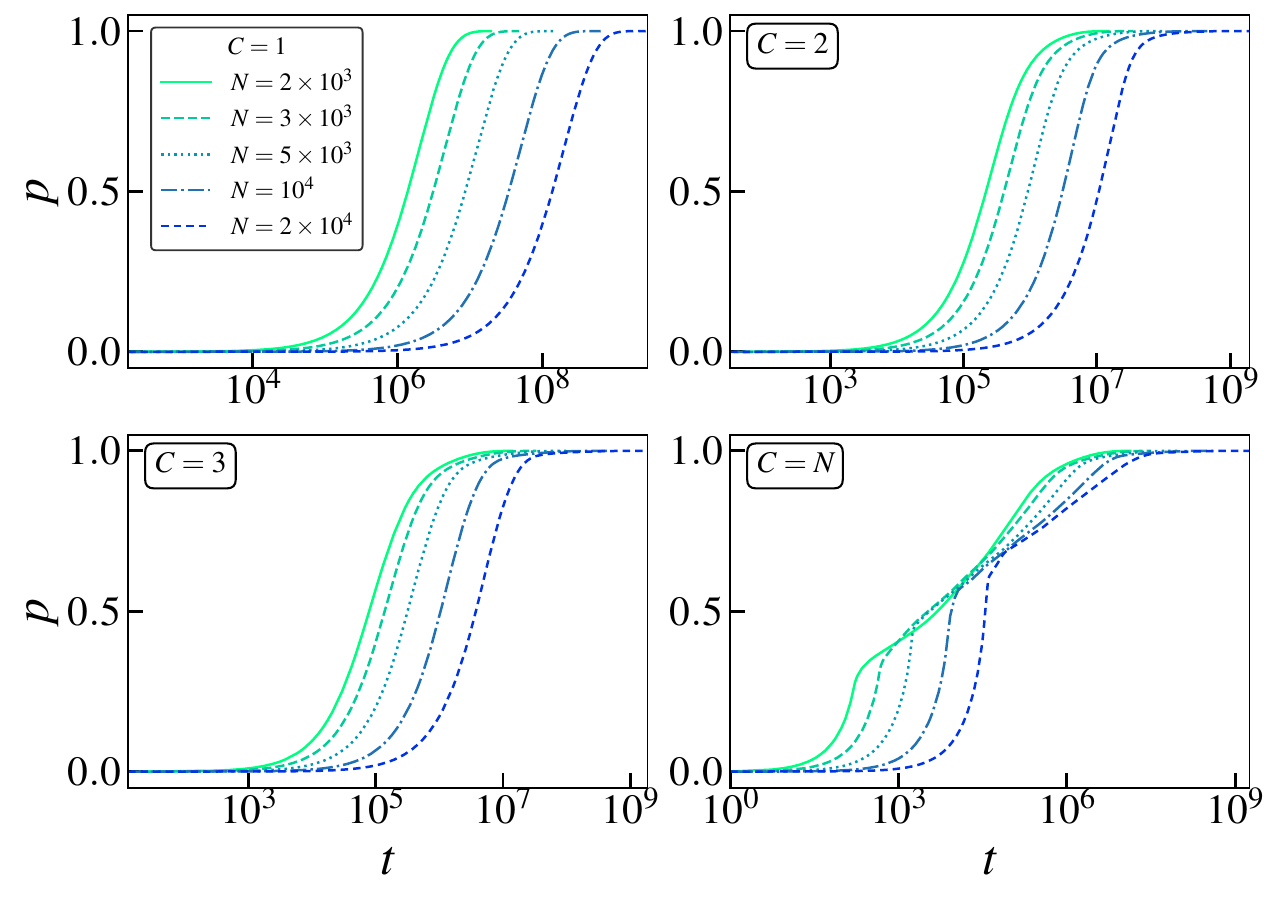}
  \caption{Fraction of removed edges $p$ [Eq.~\eqref{paramter_p}] as a function of the number of origin-destination requests $t$ for quantum internet networks under SPP dynamics with uniformly random pair selection. Panels correspond to maximum allowed shortest-path lengths $C=1$ (top left), $C=2$ (top right), $C=3$ (bottom left), and the unrestricted case $C=N$ (bottom right). The curves represent $N=2\times10^{3}$, $3\times10^{3}$, $5\times10^{3}$, $10^{4}$, and $2\times10^{4}$ nodes in a fixed spatial domain and are averaged over $10^{3}$ independent network realizations. For finite $C$, larger networks require more requests to reach the same removed-edge fraction, whereas the unrestricted dynamics exhibit a rapid initial stage followed by slower edge removal.
}
  \label{fig:demand_agent}
\end{figure}

Figure~\ref{fig:demand_agent} shows the fraction of removed edges $p$ as a function of the number of demanded node pairs $t$ for different network sizes and values of the maximum path length $C$. For the constrained dynamics $C=1$, $2$, and $3$, increasing the network size shifts the transition toward larger values of $t$, indicating that larger networks require a greater number of demanded node pairs to reach the same level of degradation. In contrast, the unrestricted case $C=N$ displays a qualitatively different evolution, characterized by a two-stage degradation process: an initial rapid increase in the fraction of removed edges, followed by a slower progression before the network becomes fully degraded. This behavior indicates that removing the path-length constraint modifies how the network degradation evolves throughout the process.

To compare the evolution of the degradation level $p(t)$ as a function of the number of demands $t$ for networks of different sizes, we rescale the control variable to $t \mapsto  tN^{-\theta}$, where $\theta$ is a scaling exponent that characterizes how the scale of the fragmentation process grows with the network size. The value of $\theta$ is determined by enforcing the best possible collapse of the curves corresponding to different values of $N$ when plotted as a function of the rescaled variable $tN^{-\theta}$. The optimal value of $\theta$ is defined as the one that minimizes the dispersion among the rescaled curves.

\begin{figure}[!htbp]
  \centering
  \includegraphics[width=\linewidth]{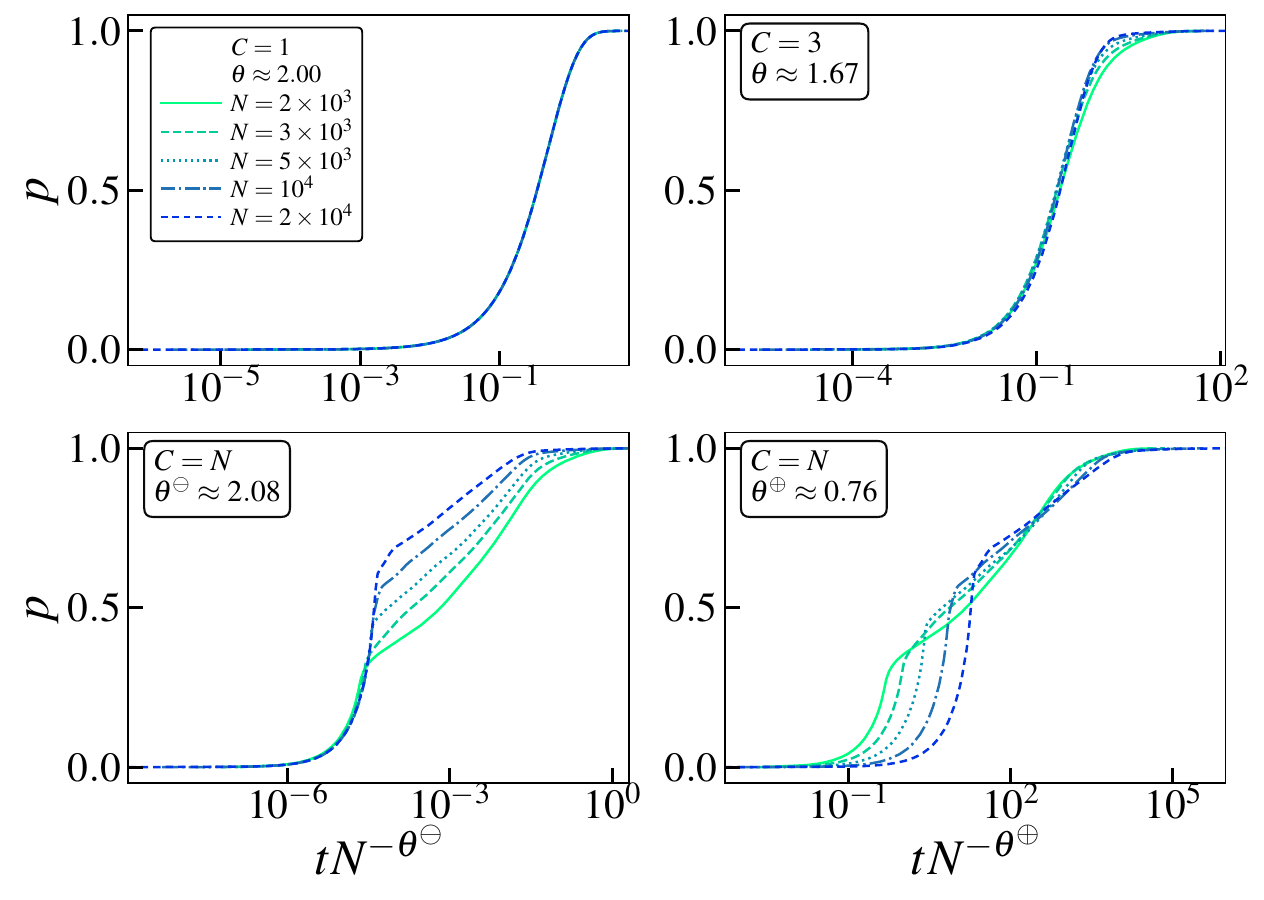}
  \caption{Finite-size rescaling of the removed-edge fraction $p(t,N)$ for quantum internet networks in a fixed spatial domain under SPP dynamics with uniformly random pair selection. The upper panels show $p$ as a function of $tN^{-\theta}$ for $C=1$ (left) and $C=3$ (right), yielding a satisfactory collapse with $\theta\approx2.00$ and $1.67$, respectively. The lower panels show the unrestricted case $C=N$, using $\theta^{\ominus}\approx2.08$ (left) and $\theta^{\oplus}\approx0.76$ (right) to describe the regimes before and after the pseudocritical threshold $t_c(N)$, respectively. These rescalings bring the curves for different network sizes closer together, but residual size-dependent fluctuations remain, particularly after the transition. Thus, the unrestricted dynamics do not exhibit a complete data collapse under the proposed scaling forms.}
  \label{fig:collapse_C1_C2_C3_CN}
\end{figure}

For finite costs $C=1, 2$ and $3$, a single exponent $\theta$ is sufficient to describe the collapse over the whole process. In these cases, a robust collapse of the rescaled curves is observed (see Fig.~\ref{fig:collapse_C1_C2_C3_CN}), although small residual dispersions become visible for the largest network sizes. In the case of infinite cost, the fragmentation dynamics exhibits two distinct temporal behaviors, associated with the phases before and after macroscopic fragmentation. For this reason, we estimate separately the exponents $\theta^{\ominus}$ and $\theta^{\oplus}$ as shown in Fig.~\ref{fig:collapse_C1_C2_C3_CN}. As the network size increases, the quality of the collapse deteriorates, indicating that the proposed rescalings do not fully superpose the curves over the entire unrestricted process.

\begin{figure}[!htbp]
  \centering
\includegraphics[width=\linewidth]{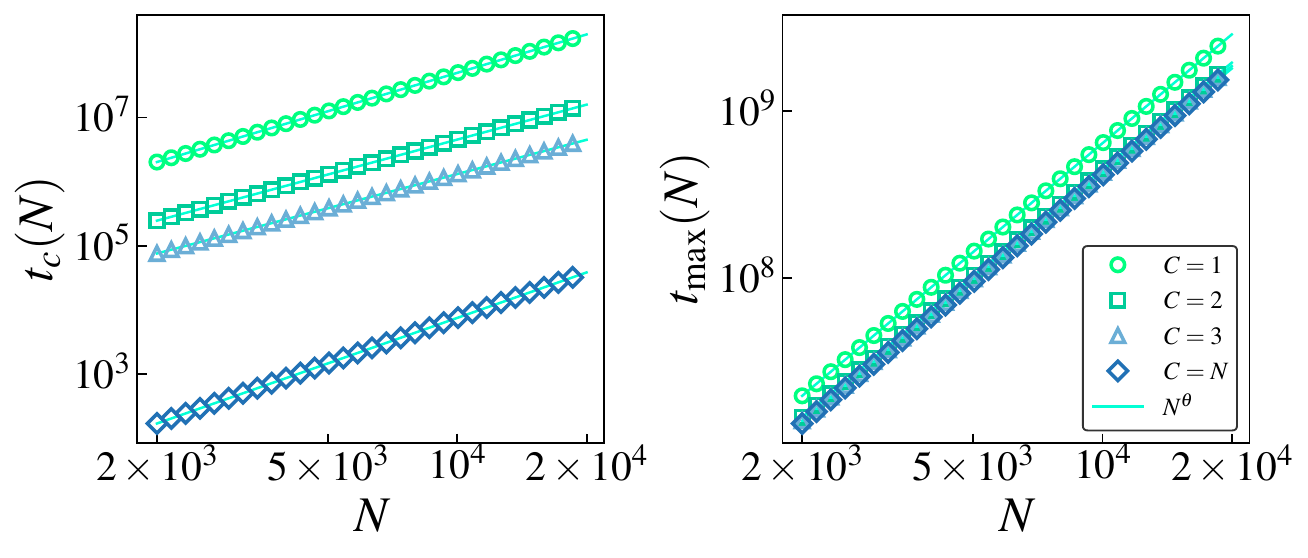}
 \caption{Pseudocritical time $t_c(N)$ (left) and complete dismantling time $t_{\max}(N)$ (right) as functions of network size $N$ for quantum internet networks in a fixed spatial domain under SPP dynamics with uniformly random pair selection. Both times count origin--destination requests: $t_c$ corresponds to the maximum of $S$, while $t_{\max}$ marks the point at which no removable shortest paths remain. Results are shown for $C=1$, $2$, $3$, and $N$, with solid lines indicating power-law fits. The fitted exponent for $t_c$ depends on $C$, whereas $t_{\max}$ grows approximately quadratically over the investigated range of system sizes, with little dependence on the path-length constraint. All fits yield coefficients of determination $r^2\approx1.00$.} \label{fig:t_critic}
\end{figure}

As shown in Fig.~\ref{fig:t_critic}, both the pseudocritical time, $t_c$, corresponding to the maximum of the susceptibility, and the complete dismantling time, $t_{\max}$, at which no removable shortest paths remain, are well described by power-law fits over the investigated range of system sizes, with exponents estimated separately for each observable and each value of $C$. While the exponent associated with $t_c(N)$ depends on the path-length constraint $C$, varying from approximately $1.67$ to $2.1$, the scaling of $t_{\max}(N)$ is nearly independent of $C$. In particular, for the representative case $C=1$, we obtain $\theta\approx2.15$, which accurately characterizes the behavior observed for all values of $C$. For $t_{\max}(N)$, we also considered logarithmically corrected fits of the form $A N^{\theta}\log N$, where $A$ is a fitted prefactor. These fits yield $\theta\approx2.05$, $2.01$, $2.01$, and $2.01$ for $C=1$, $2$, $3$, and $N$, respectively. Although the logarithmically corrected fits provide a marginal improvement for the finite-cost cases, both fitting forms are practically indistinguishable over the range of system sizes considered. In all cases, the corresponding coefficients of determination are $r^{2}\approx 1$, indicating excellent agreement between the numerical data and the proposed scaling laws. Therefore, for simplicity, we present only the pure power-law fits for $t_c(N)$ and $t_{\max}(N)$, interpreting the fitted exponents as effective descriptions over the investigated range of system sizes.

In quantum networks, when a large connected component of entangled links is present, entanglement requests between most pairs of nodes can be successfully routed through short paths. At early times, each demanded origin-destination pair consumes only a small number of links, since the shortest available paths are typically short. As the network progressively depletes, the successful paths become longer and less frequent, increasing the number of links consumed per successful demand. The critical regime corresponds to the rapid consumption of quantum links caused by a small fraction of requested node pairs. In this regime, the network loses its large-scale connectivity and fragments into several disconnected components. Following this extensive consumption of quantum links, two randomly chosen nodes are unlikely to belong to the same connected component, strongly reducing the probability of finding an available entangled path between them. Consequently, the rate at which links are consumed per demanded pair decreases sharply.

Our results for the SPP model are comparable to those obtained in Ref.~\cite{Kim_2024}. For $C=1$, we find $\theta \approx 2.00$, equal to the value obtained for ER networks in Ref.~\cite{Kim_2024}. In the infinite-cost case, we identify two distinct behaviors, characterized by $\theta^{\ominus} \approx 2.08$ and $\theta^{\oplus} \approx 0.76$, whereas \cite{Kim_2024} reports $\theta^{\ominus} \approx 0.86$ and $\theta^{\oplus} \approx 2.07$ for ER networks. Thus, although both studies reveal two characteristic scaling exponents in the infinite-cost regime, the exponents are associated with different behaviors. Moreover, in the present work, the curves corresponding to different system sizes do not exhibit a complete collapse under the proposed rescaling over the investigated range. This behavior qualitatively differs from that reported in \cite{Kim_2024}, where curve collapse is observed beyond a certain time $t$ for both $\theta^{\ominus}$ and $\theta^{\oplus}$, even in the infinite-cost regime.

\subsection{Sharpness of the percolation transition}

Here we quantify the sharpness of the transition in the SPP process through the width of the transition window \cite{explosion, Nagler_2011}, defined as $\Delta p = p_1 - p_2$, where $p_2$ denotes the largest value of the control parameter for which the largest connected component still satisfies $P > 0.5$, and $p_1$ is the largest value for which $P > 1/\sqrt{N}$. In an analogous way, we define the width of the transition in terms of the number of demanded origin-destination pairs as $\Delta \tilde{t} = (t_1 - t_2) N^{-2}$.

\begin{figure}[!htbp]
  \centering
  \includegraphics[width=\linewidth]{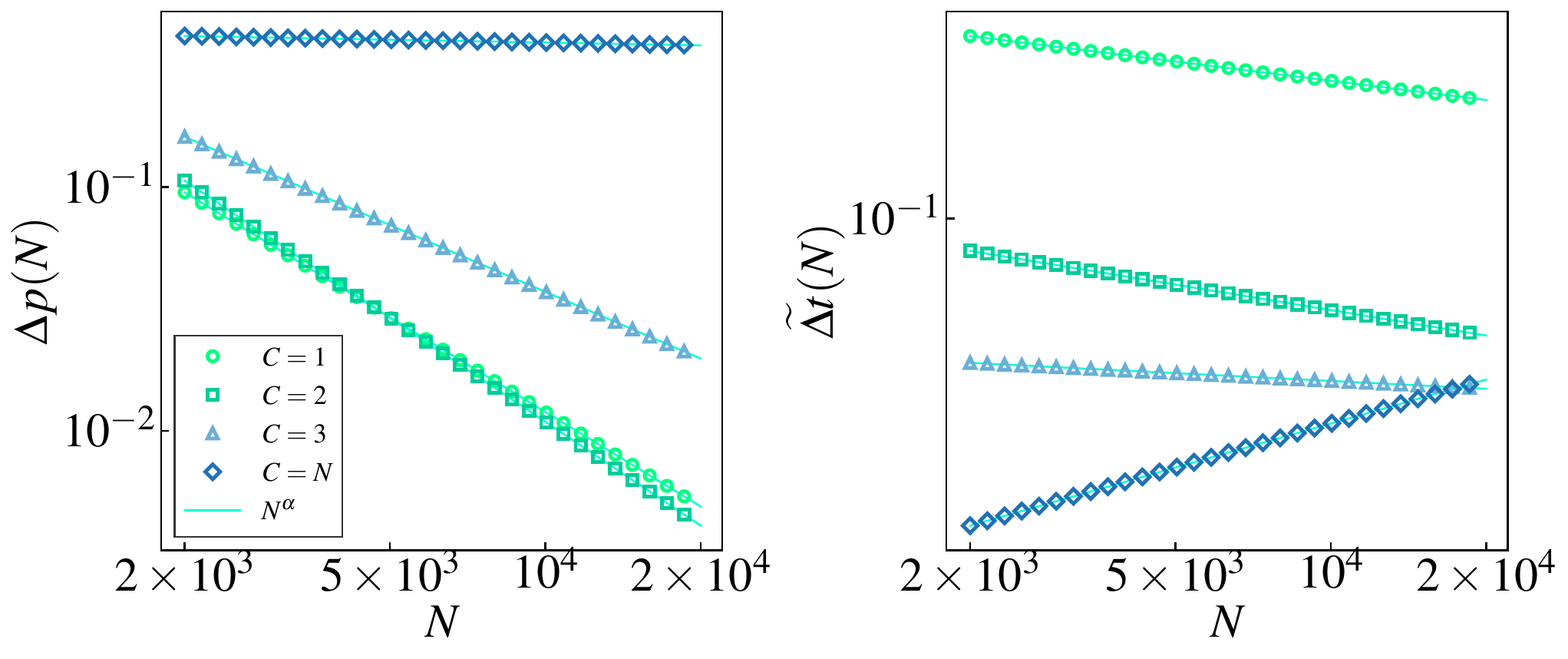}
  \caption{Scaling of the transition widths $\Delta p(N)$ and 
  $\widetilde{\Delta t}(N)$ as functions of the system size $N$ for different 
  values of the cost parameter $C$. The symbols correspond to the numerical 
  results, while the solid lines represent the power-law fits 
  $\Delta p(N)\sim N^{-\alpha}$ and 
  $\widetilde{\Delta t}(N)\sim N^{-\alpha'}$, as reported in 
  Tables~\ref{table:janela_p} and~\ref{tab:alpha_ttilde}. For finite values of 
  $C$, both transition widths decrease as the system size increases, indicating 
  that the transition becomes progressively sharper. In contrast, for $C=N$, 
  $\Delta p(N)$ remains approximately constant, whereas 
  $\widetilde{\Delta t}(N)$ increases with $N$, highlighting the distinct 
  scaling behavior of the unrestricted case.}
  \label{fig:window_transition}
\end{figure}

These quantities measure the extent of the drop in $P$ between the prescribed levels, expressed in terms of either $p$ or the normalized request count $tN^{-2}$. Their dependence on $N$ provides a finite-size measure of how sharply this degradation occurs in the chosen variables~\cite{Kim_2024}. The scaling behavior of the transition widths $\Delta p(N)$ and $\Delta \tilde{t}(N)$ is shown in Fig.~\ref{fig:window_transition}. The corresponding power-law exponents and coefficients of determination obtained from the fits are reported in Tables~\ref{table:janela_p} and \ref{tab:alpha_ttilde}.

\begin{table}[b!]
\centering
\setlength{\tabcolsep}{22pt}   
\renewcommand{\arraystretch}{1.3} 
\begin{tabular}{c c c}
\hline\hline
$C$ & $\alpha$ & $r^2$ \\
\hline
$1$      & $1.289 \pm 0.042$ & $1.00$ \\
$2$      & $1.416 \pm 0.032$ & $1.00$ \\
$3$      & $0.910 \pm 0.017$ & $1.00$ \\
$\infty$ & $0.038 \pm 0.035$ & $0.29$ \\
\hline\hline
\end{tabular}
\caption{Scaling exponents $\alpha$ obtained from fits of the transition width in the removed-edge fraction, $\Delta p\sim N^{-\alpha}$, for quantum internet networks in a fixed spatial domain at different maximum allowed shortest-path lengths $C$. The fits use linear regression in logarithmic coordinates. The quoted uncertainties are the standard errors of the fitted exponents, and $r^2$ is the corresponding coefficient of determination. The unrestricted case $C \rightarrow \infty$ is implemented using $C=N$.}\label{table:janela_p}
\end{table}

\begin{figure}[!htbp]
  \centering
  \includegraphics[width=\linewidth]{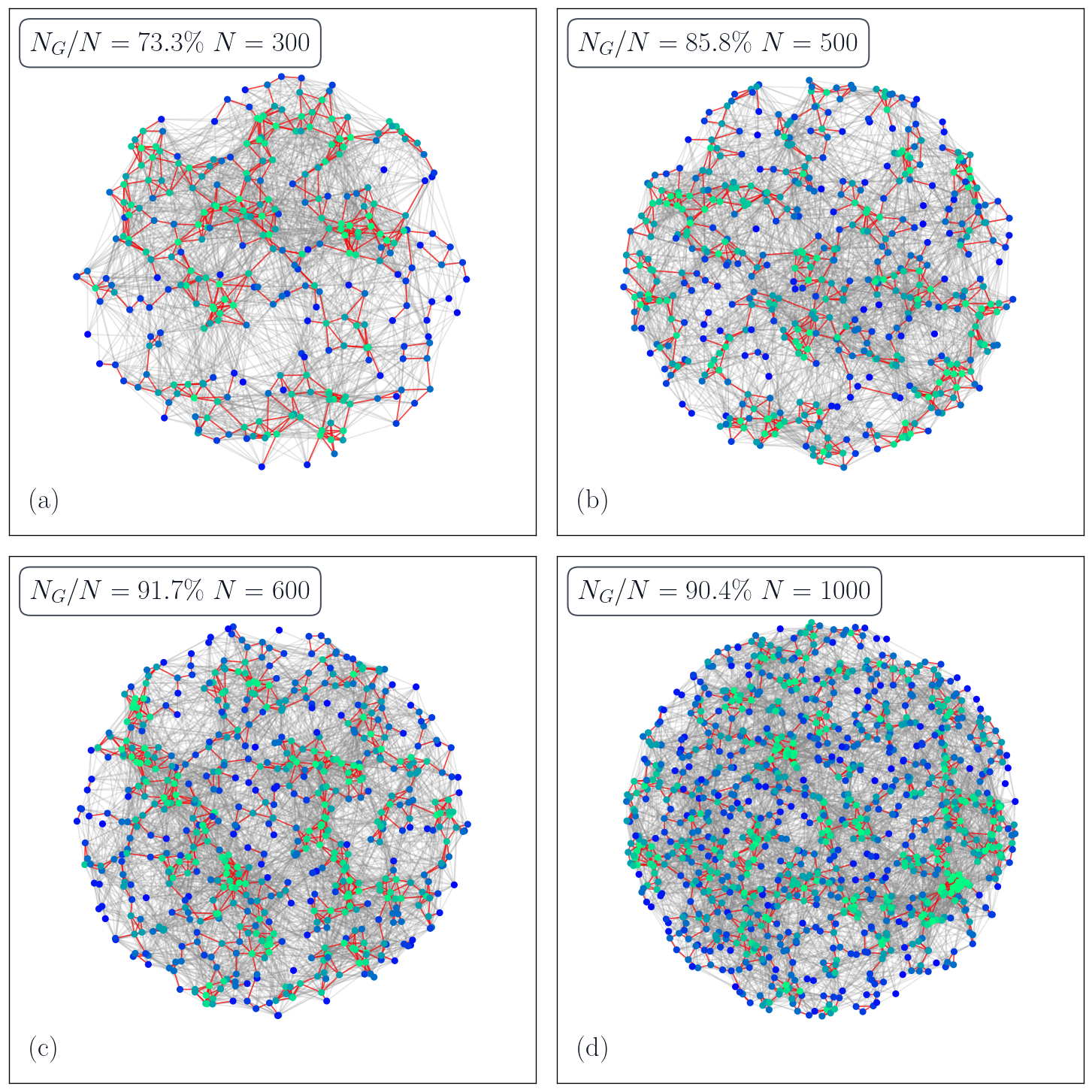}
  \caption{Illustrations of quantum internet networks with approximately constant average photonic degree $\langle k\rangle\approx4$. Gray edges represent the underlying fiber-optic infrastructure, while red edges indicate the photonic links established through photon transmission. Nodes are distributed uniformly over a disk whose radius increases as $R(N)\propto N^{1/2}$, maintaining approximately constant node density as the network grows. This construction allows the system size to increase without systematically increasing the number of photonic connections per node.}
\label{fig:quantum_network_contante}
\end{figure}

We characterize the size dependence of these widths through the fits $\Delta p\sim N^{-\alpha}$ and $\Delta\tilde{t}\sim N^{-\alpha'}$. A positive fitted exponent describes a narrowing of the corresponding interval over the investigated range, an exponent close to zero indicates weak size dependence, and a negative exponent describes broadening. It is important to note that a narrowing observed over a finite range of system sizes does not, by itself, establish a discontinuous transition in the thermodynamic limit~\cite{RiordanWarnke2012}.

The estimation of the scaling exponents is performed by means of log-log fits. Taking the logarithm of both sides of the relations above, we obtain
\begin{equation}
\begin{aligned}
\log \Delta p(N) &= -\alpha \log N + c_{1}, \\
\log \widetilde{\Delta t}(N) &= -\alpha' \log N + c_{2},
\end{aligned}
\label{eq:transition-width-scaling}
\end{equation}
where $c_{1}$ and $c_{2}$ are constants. The exponents and their standard errors are estimated from the linear regression fit, while the quality of the fit is assessed through the coefficient of determination $r^2$.

\begin{table}[b!]
\centering
\setlength{\tabcolsep}{22pt}   
\renewcommand{\arraystretch}{1.3} 
\begin{tabular}{c c c}
\hline\hline
$C$ & $\alpha'$ & $r^2$ \\
\hline
$1$      & $0.186 \pm 0.025$  & $0.95$ \\
$2$      & $0.246 \pm 0.014$  & $1.00$ \\
$3$      & $0.075 \pm 0.016$  & $0.89$ \\
$\infty$ & $-0.425 \pm 0.077$ & $0.91$ \\
\hline\hline
\end{tabular}
\caption{Scaling exponents $\alpha'$ obtained from fits of the normalized transition width, $\Delta\tilde{t}\sim N^{-\alpha'}$, for quantum internet networks in a fixed spatial domain at different maximum allowed shortest-path lengths $C$. The fits use linear regression in logarithmic coordinates. The quoted uncertainties are the standard errors of the fitted exponents, and $r^2$ is the corresponding coefficient of determination. The unrestricted case $C \rightarrow \infty$ is implemented using $C=N$.}\label{tab:alpha_ttilde}
\end{table}

For ER networks, in Ref.~\cite{Kim_2024} it was shown that the transition widths follow the scaling laws $\Delta p \sim N^{-\alpha}$ and $\Delta \widetilde{t} \sim N^{-\alpha'}$. For finite costs $C$, the corresponding exponents are $\alpha = 0.02 \pm 0.01$, $0.03 \pm 0.01$, and $0.03 \pm 0.01$, while $\alpha' = 0.03 \pm 0.01$, $0.03 \pm 0.01$, and $0.04 \pm 0.01$, respectively. These values indicate a slow reduction of both transition widths as the system size increases. In contrast, for the quantum networks considered in this work, significantly larger exponents are obtained for finite values of $C$, namely $\alpha = 1.289$, $1.416$, and $0.910$, and $\alpha' = 0.186$, $0.246$, and $0.075$, for $C=1$, $2$, and $3$, respectively. These results show that both transition widths decrease much more rapidly with increasing $N$, indicating that the percolation transition becomes significantly more abrupt than in ER networks.

For the unrestricted case, however, the observed behavior differs qualitatively. While \cite{Kim_2024} finds $\alpha=0.13\pm0.01$ and $\alpha'=0.88\pm0.01$, corresponding to decreasing transition widths, our results for quantum networks show no clear narrowing over the investigated range of system sizes. In particular, $\Delta p(N)$ varies only weakly with $N$, and its power-law fit has a low coefficient of determination ($r^2\approx0.29$), whereas $\Delta\tilde{t}(N)$ increases with $N$. These observations characterize the finite-size behavior without establishing the asymptotic limits of the widths.

\begin{figure*}[!tp]
  \centering
\includegraphics[width=\textwidth]{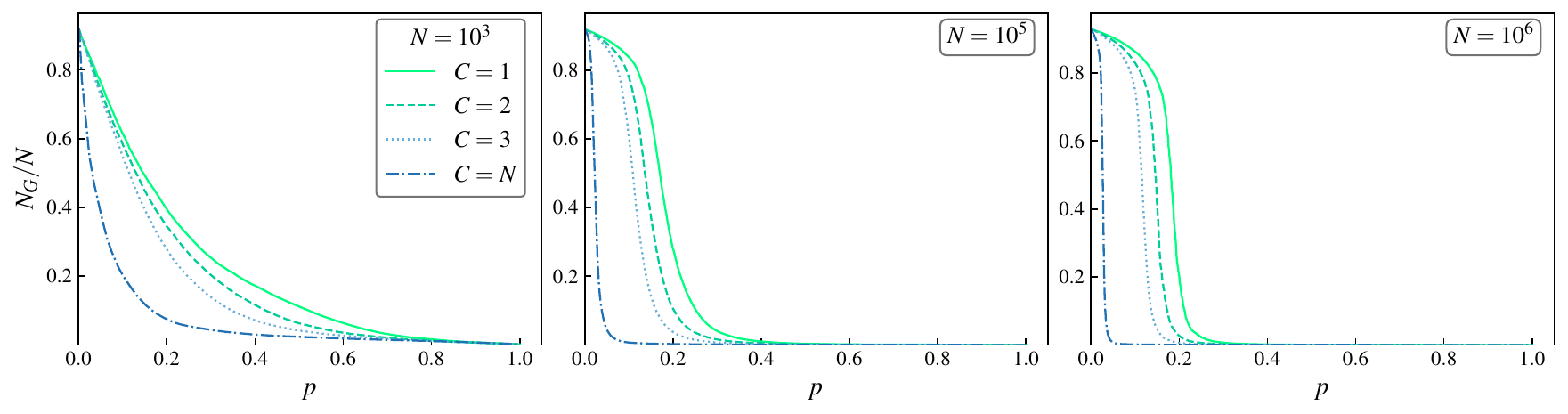}
 \caption{Relative size of the giant component, $N_G/N$, as a function of the fraction of removed edges $p$ [Eq.~\eqref{paramter_p}] for quantum internet networks under SPP dynamics with uniformly random origin--destination pair selection. From left to right, the panels correspond to $N=10^{3}$, $10^{5}$, and $10^{6}$ nodes. The radius of the spatial domain is rescaled as $R(N)\propto N^{1/2}$ to maintain an approximately constant initial average degree $\langle k\rangle\approx4$. The curves correspond to maximum allowed shortest-path lengths $C=1$, $2$, $3$, and $N$, with $C=N$ denoting the unrestricted dynamics. Each curve represents an average over $10^{3}$ independent network realizations.}
 \label{fig:spp-quantum-cluster-random-constante}
\end{figure*}

Furthermore, the location of the degradation and the width of the interval over which it occurs describe different aspects of the dynamics. As shown in Fig.~\ref{fig:spp-quantum-cluster-random}, the largest connected component reaches low values at smaller values of the removed-edge fraction $p$ for $C=N$ than for the finite costs considered. Nevertheless, $\Delta p(N)$ shows no clear narrowing over the investigated range of system sizes. This behavior can be understood as a consequence of the spatial embedding of the quantum network. Unlike ER networks, whose connectivity is independent of the physical positions of the nodes, the quantum network is embedded in a geometric domain where shortest paths are strongly influenced by spatial distance. When $C=N$, the absence of a fixed path-length cutoff permits the removal of both short and long paths. This broader range of path lengths may contribute to the observed differences from ER networks, although the present results do not establish whether a single characteristic length governs the fragmentation process.

\subsection{Analysis of Quantum Networks with Controlled Mean Degree}

Let us now analyze quantum networks generated with a fixed mean degree. In the quantum internet, the average degree is directly related to the number of communication links and the routing alternatives available for entanglement distribution. By keeping the initial average degree approximately constant, we study how network fragmentation depends on system size without a systematic increase in the mean number of photonic links per node. Other structural properties, including characteristic shortest-path lengths, may still vary with $N$.

The average degree $\langle k \rangle$ of quantum networks has a linear relationship with the density $\rho$, given by $\langle k \rangle = A \rho$, where $A = 5.2 \times 10^{4}$ is a constant independent of the network size $N$. To keep the average degree constant, we first note that the density is defined as $\rho = N/\pi R^{2}$, where $R>0$ is the radius of the disk in which the nodes are placed. If $R$ is kept fixed, the density increases as $N$ grows, which changes the average degree.

Using the relation $\rho = \langle k \rangle / A$, we obtain an explicit expression for the radius:
\begin{align}
R = \sqrt{\frac{A N}{\langle k \rangle \pi}}.
\end{align}
Therefore, the radius must be treated as a function of $N$, that is, $R = R(N)\sim \sqrt{N}$. By adjusting $R$ as $N$ increases, we keep $\rho$ and consequently the average degree constant. Figure~\ref{fig:quantum_network_contante} shows examples of quantum internet networks generated with approximately constant average degree.

As in the previous cases, we consider a process in which both nodes of each selected pair are chosen uniformly at random. Under this criterion, we analyze the emergence of the giant component, as shown in Fig.~\ref{fig:spp-quantum-cluster-random-constante}. As we see in the figure, as $N$ increases, the giant component collapses at progressively smaller values of $p$. Increasing $C$ further accelerates the fragmentation process, with $C=N$ yielding the earliest collapse.

\begin{figure}[!b]
  \centering
\includegraphics[width=\linewidth]{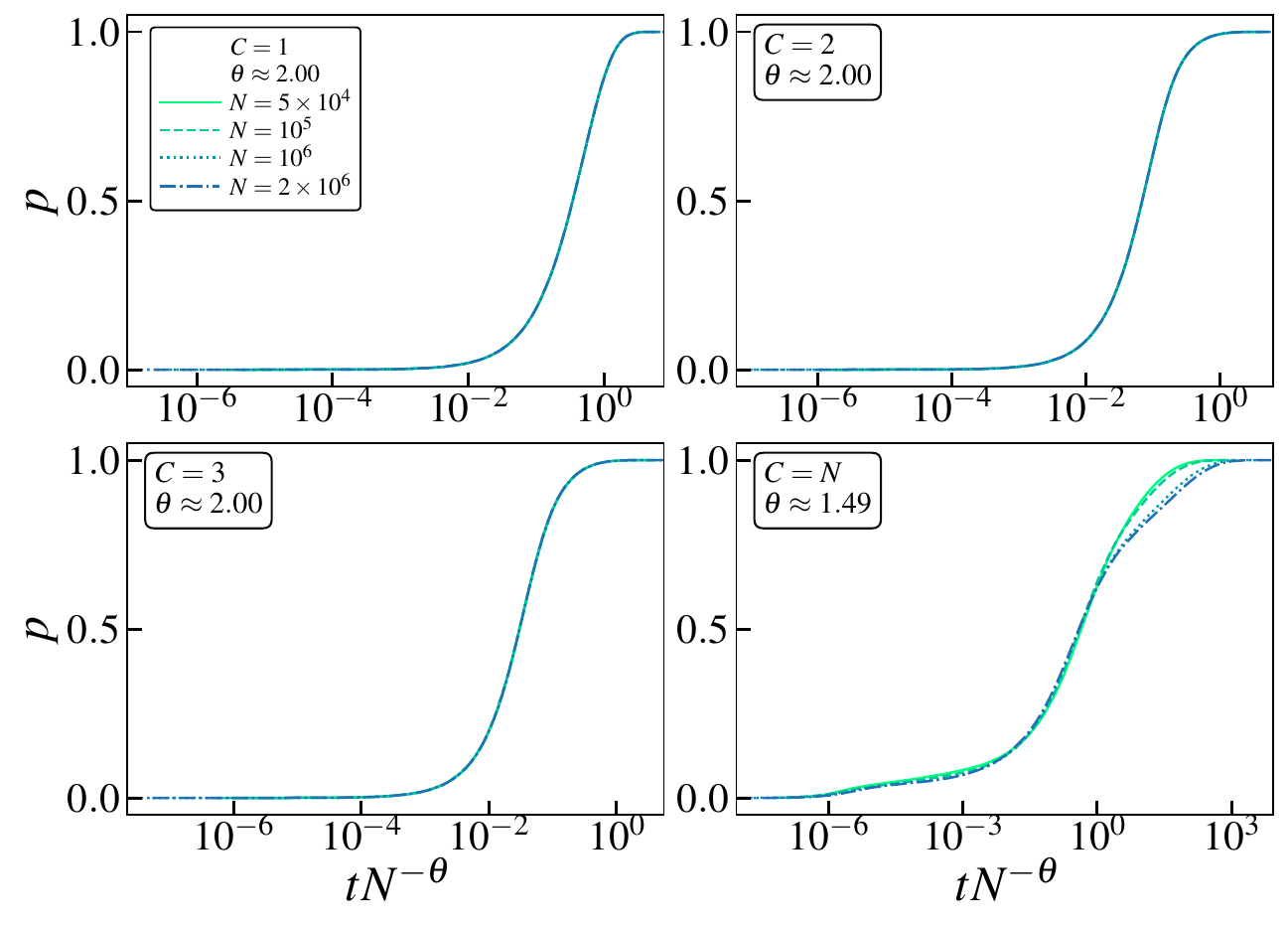}
 \caption{Fraction of removed edges $p(t,N)$ versus the rescaled request count $tN^{-\theta}$ for quantum internet networks with approximately constant initial average degree $\langle k\rangle\approx4$ under SPP dynamics with uniformly random pair selection. Panels correspond to $C=1$ (top left), $C=2$ (top right), $C=3$ (bottom left), and $C=N$ (bottom right). The curves represent $N=5\times10^{4}$, $10^{5}$, $10^{6}$, and $2\times10^{6}$ nodes. For $C=1$, $2$, and $3$, $\theta\approx2$ yields a clear data collapse. For the unrestricted case $C=N$, the optimal exponent $\theta\approx1.49$ brings the curves closer together, but noticeable deviations persist near the transition, indicating incomplete collapse under the proposed rescaling.}
 \label{fig:colapse_contante}
\end{figure}

For the unrestricted case $C=N$, the exponent $\theta\approx1.49$ brings the curves corresponding to different system sizes closer together, but deviations remain near the transition, as shown in Fig.~\ref{fig:colapse_contante}. In contrast, the finite-cost cases exhibit a clear collapse under the proposed rescaling. The broader range of admissible path lengths in the unrestricted case may contribute to its more complex temporal evolution. However, the incomplete collapse could also reflect finite-size corrections or the need for a different scaling form. The present results therefore indicate limitations of the proposed rescaling over the investigated range of system sizes.

There are clear differences between the cases in which the mean degree is not kept constant and the case in which one imposes $\langle k \rangle \approx 4$. In the first case, where nodes are distributed within a fixed-radius spatial domain and the node density increases with $N$, large scaling exponents are observed in the fit $\Delta p \sim N^{-\alpha}$ (see Table~\ref{table:janela_p}). This behavior indicates a rapid narrowing of the transition window as the network size increases, leading to a progressively sharper degradation of the global connectivity. On the other hand, when the mean degree is kept constant by rescaling the domain as $R \sim \sqrt{N}$, the exponents $\alpha$ associated with $\Delta p$ become significantly smaller for finite costs (see Table~\ref{tab:alpha_dp_const}), indicating that the width of the transition window decreases much more slowly with $N$.

\begin{figure}[!htbp]
  \centering
\includegraphics[width=\linewidth]{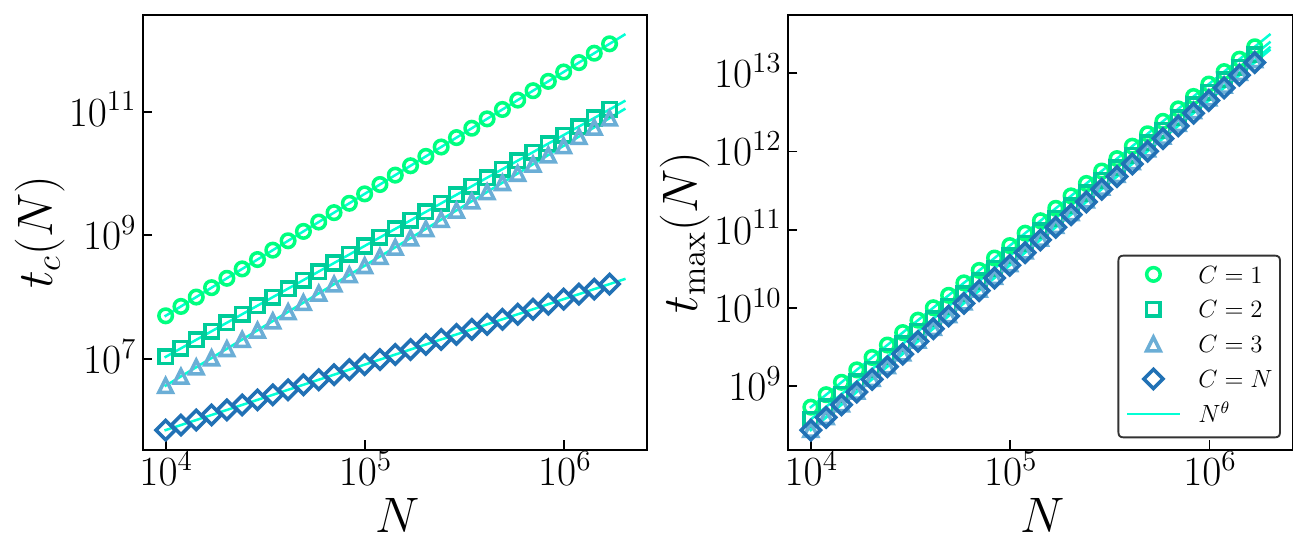}
 \caption{Pseudocritical time $t_c(N)$ and the number of demanded pairs required to completely dismantle the network, $t_{\max}(N)$, as functions of the system size $N$ for different maximum shortest-path lengths $C$, considering photonic networks with approximately constant initial average degree $\langle k\rangle\approx4$. The solid lines represent power-law fits, with exponents estimated separately for $t_c(N)$ and $t_{\max}(N)$ at each value of $C$. While the scaling of $t_c(N)$ depends on $C$, particularly in the unrestricted case $C=N$, the complete dismantling time $t_{\max}(N)$ grows approximately quadratically over the investigated range of system sizes, with little dependence on the path-length constraint.}
\label{fig:t_critic_constant}
\end{figure}

For finite costs, both network construction schemes yield characteristic times that increase with system size, as shown in Fig.~\ref{fig:t_critic_constant}. At approximately constant average degree, the pseudocritical time has fitted exponents $\theta\approx1.98$, $1.87$, and $1.94$ for $C=1$, $2$, and $3$, respectively, indicating growth close to quadratic over the investigated range. Likewise, the right panel indicates that power-law fits to $t_{\max}(N)$ yield similar effective exponents for all finite costs under both network construction schemes over the investigated range of system sizes.

A different behavior is observed for the unrestricted case $C=N$. While the effective exponent obtained from power-law fits to $t_{\max}(N)$ remains approximately unchanged over the investigated range, the exponent associated with the pseudocritical threshold decreases from $\theta\approx2.08$, obtained when the average degree increases with $N$, to $\theta\approx1.16$ when the average degree is kept approximately constant. This result indicates that, unlike the dynamics in the finite-cost cases, the unrestricted dynamics is highly sensitive to the evolution of the network's global connectivity. As the average degree increases with the system size, the network becomes progressively more connected, requiring a larger number of pair demands to produce the same level of connectivity degradation, which leads to a stronger growth of $t_c(N)$. By contrast, when the average degree remains approximately constant, this increase in connectivity no longer occurs as the system grows, making the degradation process more efficient and resulting in a substantially smaller scaling exponent.

For finite costs, $\alpha'$ remains positive under both network construction schemes, although its fitted value depends on the construction, particularly for $C=2$ and $3$, as shown in Tables~\ref{tab:alpha_ttilde} and \ref{tab:alpha_ttilde_const}. However, the case $C=N$ exhibits a qualitatively different behavior. When the mean degree is not kept constant, the fitted exponent is negative ($\alpha'=-0.425$, see Table~\ref{tab:alpha_ttilde}), indicating that the normalized transition width increases with $N$ over the investigated range. In contrast, maintaining the mean degree approximately constant yields $\alpha'\approx 0.97$ (see Table~\ref{tab:alpha_ttilde_const}), indicating that this width decreases with increasing system size.

\begin{figure}[!t]
  \centering
\includegraphics[width=\linewidth]{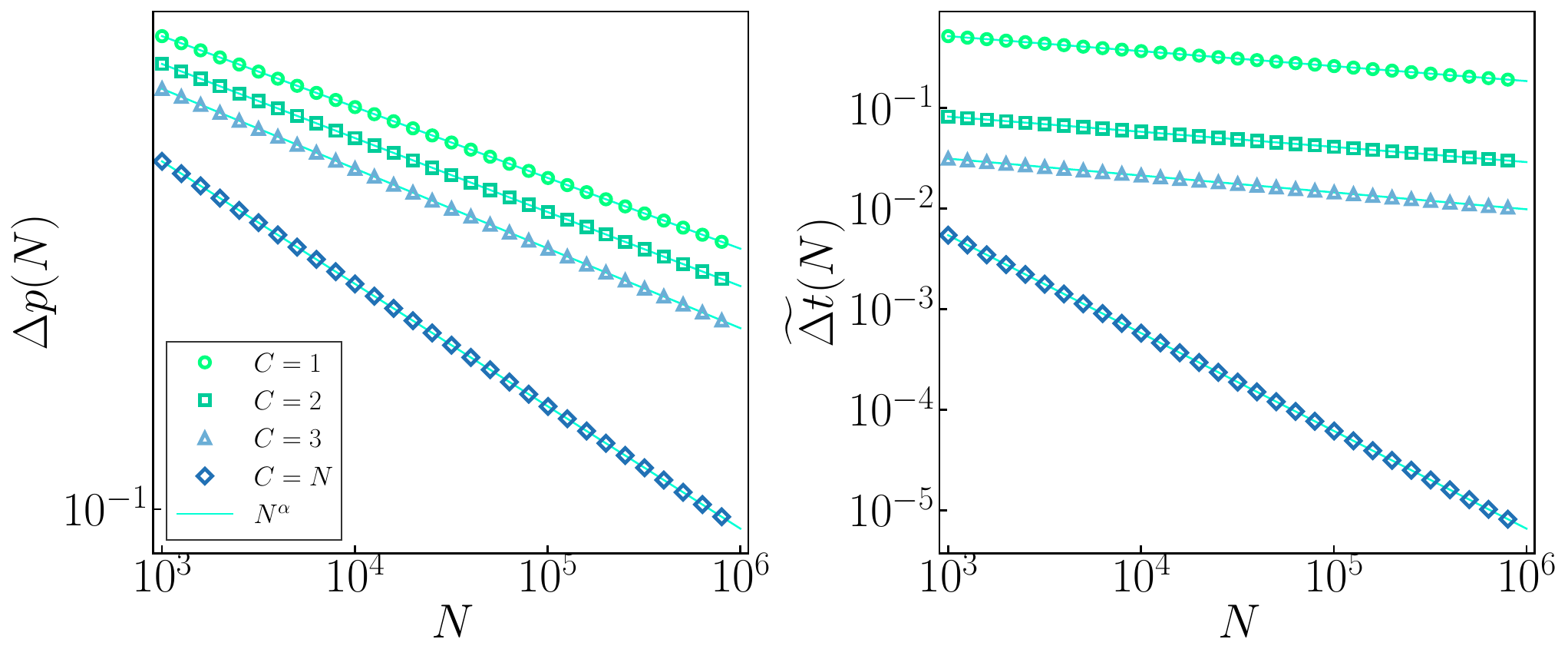}
\caption{Scaling of the transition window, $\Delta p(N)$, and its corresponding normalized width in terms of the number of demanded pairs, $\widetilde{\Delta t}(N)$, as functions of the system size $N$ for quantum internet networks with approximately constant initial average degree $\langle k\rangle\approx4$ and different maximum allowed shortest-path lengths $C$. The symbols represent the numerical results, while the solid lines correspond to the power-law fits $\Delta p(N)\sim N^{-\alpha}$ and $\widetilde{\Delta t}(N)\sim N^{-\alpha'}$. The fitted exponents and coefficients of determination are reported in Tables~\ref{tab:alpha_dp_const} and~\ref{tab:alpha_ttilde_const}. Both widths decrease as the system size increases, with the unrestricted case $C=N$ exhibiting a substantially faster decay and a more pronounced narrowing of the transition region.}
\label{fig:window_transition_constant}
\end{figure}

The finite-size behavior of the transition windows is also modified when the mean degree is kept constant. Both $\Delta p$ and $\Delta\tilde{t}$ decrease monotonically with the system size for all values of $C$, as shown in Fig.~\ref{fig:window_transition_constant}, while the unrestricted case exhibits the fastest decay. This behavior arises because keeping the mean degree constant prevents the network from becoming progressively more connected as $N$ increases. Consequently, the unrestricted dynamics becomes more effective at degrading the network connectivity, producing a faster contraction of the transition region. Thus, the transition becomes progressively sharper in the chosen variables over the investigated range of system sizes.

\begin{table}[b!]
\centering
\setlength{\tabcolsep}{22pt}
\renewcommand{\arraystretch}{1.3}
\begin{tabular}{c c c}
\hline\hline
$C$ & $\alpha$ & $r^2$ \\
\hline
$1$      & $0.11\pm 0.008$ & $1.00$ \\
$2$      & $0.11\pm 0.008$ & $1.00$ \\
$3$      & $0.12\pm 0.007$ & $1.00$ \\
$\infty$      & $0.20\pm 0.005$ & $1.00$ \\
\hline\hline
\end{tabular}
\caption{Scaling exponents $\alpha$ obtained from fits of the transition width in the removed-edge fraction, $\Delta p\sim N^{-\alpha}$, for quantum internet networks with approximately constant initial average degree $\langle k\rangle\approx4$ at different maximum allowed shortest-path lengths $C$. The fits use linear regression in logarithmic coordinates. The quoted uncertainties are the standard errors of the fitted exponents, and $r^2$ is the corresponding coefficient of determination. The unrestricted case $C\rightarrow\infty$ is implemented using $C=N$.}\label{tab:alpha_dp_const}
\end{table}

In summary, keeping the average degree approximately constant modifies the fitted scaling of the transition widths for both finite and unrestricted costs. In the unrestricted case, this change is particularly evident in the normalized transition width, which increases with $N$ for a fixed spatial domain but decreases when the average degree is kept approximately constant. Since the average number of physical links per node remains constant as $N$ increases, network growth does not increase local connectivity. For finite costs, the scaling exponents of $t_c(N)$ remain close to $2$, in agreement with those reported by Ref.~\cite{Kim_2024}. In contrast, for $C=N$, the exponent of $t_c(N)$ decreases from approximately $2.08$ to $1.16$, while the normalized transition width scales as $\Delta\tilde{t}\sim N^{-0.97}$. These results indicate that, when there is no restriction on the length of the shortest paths that can be removed, the network reaches the critical threshold after processing an increasingly smaller fraction of all possible node pairs. In the context of a fiber-based quantum internet, this means that the links used for entanglement distribution become unavailable more rapidly, accelerating network fragmentation. Because the average degree remains constant, increasing the network size is not sufficient to offset this effect, leading to a sharper transition in the $C=N$ case.

\begin{table}[b!]
\centering
\setlength{\tabcolsep}{22pt}
\renewcommand{\arraystretch}{1.3}
\begin{tabular}{c c c}
\hline\hline
$C$ & $\alpha'$ & $r^2$ \\
\hline
$1$      & $0.15\pm 0.007$ & $1.00$ \\
$2$      & $0.15\pm 0.007$ & $1.00$ \\
$3$      & $0.17 \pm 0.007$ & $1.00$ \\
$\infty$      & $0.97 \pm 0.007$ & $1.00$ \\
\hline\hline
\end{tabular}
\caption{Scaling exponents $\alpha'$ obtained from fits of the normalized transition width, $\Delta\tilde{t}\sim N^{-\alpha'}$, for quantum internet networks with approximately constant initial average degree $\langle k\rangle\approx4$ at different maximum allowed shortest-path lengths $C$. The fits use linear regression in logarithmic coordinates. The quoted uncertainties are the standard errors of the fitted exponents, and $r^2$ is the corresponding coefficient of determination. The unrestricted case $C\rightarrow\infty$ is implemented using $C=N$.}\label{tab:alpha_ttilde_const}
\end{table}

\subsection{Equivalent cost $C^\star$}

The case $C=N$ represents the limit in which the dynamics considers all available shortest paths, regardless of their length. Although this regime yields the maximum degradation of network connectivity, its implementation can become computationally expensive for very large networks. This naturally motivates the question of whether the unrestricted dynamics can be accurately reproduced using a finite maximum path length. In this section, we introduce an equivalent cost $C^\star$, defined as the empirical $0.999$-quantile of shortest-path lengths sampled in the initial giant component. We then assess whether this finite path-length limit reproduces the macroscopic degradation observed for $C=N$ by comparing the corresponding averaged connectivity curves. We now present a procedure to estimate this value from the shortest-path length distribution and validate its ability to reproduce the unrestricted dynamics.

Consider an unweighted network
$\mathcal{G}=(\mathcal{V},\mathcal{E})$
and let
$\mathcal{C}=(\mathcal{V}_{\mathcal C},\mathcal{E}_{\mathcal C})$
denote its giant component. For two distinct nodes
$u,v\in\mathcal{V}_{\mathcal C}$,
the shortest-path distance is defined as
\begin{align}
d_{\mathcal C}(u,v)
=
\min_{L\in\mathcal{P}(u,v)} |L|,
\end{align}
where $\mathcal{P}(u,v)$ denotes the set of all paths connecting $u$ and $v$ within the giant component, and $|L|$ represents the number of edges in the path $L$. Since the giant component is connected, there exists at least one path between any pair of nodes
$u,v\in\mathcal{V}_{\mathcal C}$.
Therefore, the shortest-path distance
$d_{\mathcal C}(u,v)$
is well defined.

The exact distribution of shortest-path distances can be obtained by evaluating all unordered pairs of nodes in the giant component. The total number of such pairs is
\begin{align}
\binom{|\mathcal{V}_{\mathcal C}|}{2}
=
\frac{|\mathcal{V}_{\mathcal C}|\left(|\mathcal{V}_{\mathcal C}|-1\right)}{2},
\end{align}
which grows quadratically with the size of the giant component. Consequently, for very large networks, computing the shortest-path distance for every pair of nodes may become computationally prohibitive.

To overcome this limitation, a sample of $M>0$ distinct unordered pairs of nodes is selected,
\begin{align}
\mathcal{S}_{M}
=
\left\{
\{u_1,v_1\},
\{u_2,v_2\},
\ldots,
\{u_M,v_M\}
\right\},
\end{align}
chosen uniformly without replacement from the set
\begin{align}
\Omega_{\mathcal C}
=
\left\{
\{u,v\}:
u,v\in\mathcal{V}_{\mathcal C},
\;
u\neq v
\right\}.
\end{align}
Hence, $
M
\leq
\binom{|\mathcal{V}_{\mathcal C}|}{2}.
$
Sampling without replacement ensures that the same unordered pair is never selected more than once. Moreover, since the pairs are unordered, we have $\{u,v\}=\{v,u\}$. For each sampled pair $\{u_i,v_i\}$, the shortest-path distance is computed as
\begin{align}
\ell_i
=
d_{\mathcal C}(u_i,v_i)
=
\min_{L\in\mathcal{P}(u_i,v_i)} |L|,
\qquad
i=1,\ldots,M.
\end{align}
This yields the sample of shortest-path distances
\begin{align}
\mathcal{L}_M
=
\{\ell_1,\ell_2,\ldots,\ell_M\}.
\end{align}
This sample provides an empirical approximation of the distribution of shortest-path distances between pairs of nodes in the giant component.

The cumulative distribution function (CDF) of the shortest-path length $\ell$ is defined as
\begin{align}
F_{\ell}(x)
=
\mathbb{P}(\ell\leq x).
\end{align}
Since the exact distribution $F_{\ell}$ is generally unavailable, we estimate it through the empirical cumulative distribution function (ECDF). Given a sample of $M$ shortest-path lengths
$\ell_1,\ell_2,\ldots,\ell_M$, the ECDF is defined as
\begin{align}
\widehat{F}_{\ell}(x)
=
\frac{1}{M}
\sum_{i=1}^{M}
\mathbf{1}_{\{\ell_i\leq x\}},
\end{align}
where
\begin{align}
\mathbf{1}_{\{\ell_i\leq x\}}
=
\begin{cases}
1, & \text{if }\ell_i\leq x,\\
0, & \text{if }\ell_i>x.
\end{cases}
\end{align}

For $q\in(0,1]$, the empirical quantile of order $q$ is defined as
\begin{align}
\widehat{\mathcal Q}_{q}
=
\inf
\left\{
x:
\widehat{F}_{\ell}(x)\geq q
\right\}.
\end{align}
Since the shortest-path distances $\ell_i$ are integers, the empirical quantile can also be obtained from the order statistics.
Let
\begin{align}
\ell_{(1)}
\le
\ell_{(2)}
\le
\cdots
\le
\ell_{(M)}
\end{align}
denote the ordered shortest-path lengths (order statistics) obtained by sorting the sampled values
$\ell_1,\ell_2,\ldots,\ell_M$
from the smallest to the largest. For a prescribed coverage level
$q\in(0,1)$,
the empirical $q$-quantile is defined as the smallest observed shortest-path length for which at least a fraction $q$ of the sampled node pairs have shortest-path lengths not exceeding it. Equivalently,
\begin{align}
\widehat{\mathcal Q}_{q}
=
\ell_{(\lceil qM\rceil)},
\end{align}
where $\lceil\cdot\rceil$ denotes the ceiling function, so that $\lceil qM\rceil$ is the smallest integer greater than or equal to $qM$.

For an integer value $C\geq 1$, we define the coverage function as
\begin{align}
f(C)
=
\mathbb{P}(\ell\leq C).
\end{align}
This quantity represents the probability that two nodes selected uniformly at random from the giant component are separated by a shortest-path distance of at most $C$ edges. Its empirical estimator is defined by
\begin{align}
\widehat{f}(C)
=
\frac{1}{M}
\sum_{i=1}^{M}
\mathbf{1}_{\{\ell_i\leq C\}}.
\end{align}
Consequently,
\begin{align}
\widehat{f}(C)
=
\widehat{F}_{\ell}(C).
\end{align}
Therefore, $\widehat{f}(C)$ corresponds to the fraction of sampled node pairs whose shortest-path distance does not exceed $C$. In the context of the SPP dynamics, this quantity estimates the fraction of demands between pairs of nodes that satisfy the path-length constraint $\ell_i\leq C$.

Complementarily, we define
\begin{align}
b(C)
=
\mathbb{P}(\ell>C)
=
1-f(C).
\end{align}
The quantity $b(C)$ represents the probability that the shortest-path distance between a pair of nodes exceeds the limit $C$. In the SPP dynamics, such pairs cannot trigger the removal of a path when only shortest paths of length at most $C$ are allowed. The corresponding empirical estimator is
\begin{align}
\widehat{b}(C)
=
\frac{1}{M}
\sum_{i=1}^{M}
\mathbf{1}_{\{\ell_i>C\}},
\end{align}
and, since the events $\{\ell_i\leq C\}$ and $\{\ell_i>C\}$ are complementary, we have
\begin{align}
\widehat{b}(C)
=
1-\widehat{f}(C).
\end{align}

To determine a value of $C$ that covers a prescribed fraction of node pairs, 
we fix a target coverage level $q$. In this work, we set $q=0.999$ and define
\begin{align}
C^{\star}
=
\min
\left\{
C\in\mathbb{N}:
\widehat{f}(C)\geq 0.999
\right\}.
\end{align}
Since $\widehat{f}(C)=\widehat{F}_{\ell}(C)$, $C^\star$ is equivalently the 
empirical $0.999$-quantile of the path-length distribution,
\begin{align}
C^{\star}
=
\widehat{\mathcal Q}_{0.999}.
\end{align}
By definition, $C^\star$ is the smallest integer for which the target 
coverage is attained, and therefore
\begin{align}
\widehat{f}(C^\star)\geq 0.999.
\end{align}

Furthermore, by the minimality of $C^{\star}$, no positive integer smaller than
$C^{\star}$ satisfies this inequality. In particular, whenever
$C^{\star}>1$,
\begin{align}
\widehat{f}(C^{\star}-1)<0.999.
\end{align}
Therefore, $C^{\star}$ is the smallest integer that covers at least $99.9\%$
of the shortest-path distances observed in the sample. Consequently, the
fraction of sampled pairs with shortest-path distances exceeding $C^\star$ is
\begin{align}
\widehat{b}(C^{\star})
=
1-\widehat{f}(C^{\star})
\leq
0.001.
\end{align}
Thus, at most $0.1\%$ of the sampled node pairs have a shortest-path distance
greater than $C^{\star}$.

It is important to note that the maximum shortest-path distance obtained from
the sampling procedure,
\begin{align}
\ell_{\max}^{(\mathrm{sample})}
=
\max_{1\leq i\leq M}\ell_i,
\end{align}
represents only the largest shortest-path distance observed among the sampled
pairs. In general, this quantity does not necessarily coincide with the
diameter of the giant component,
\begin{align}
D_{\mathcal C}
=
\max_{u,v\in\mathcal{V}_{\mathcal C}}
d_{\mathcal C}(u,v),
\end{align}
since a pair of nodes attaining the diameter may not be included in the sample.
Therefore,
\begin{align}
\ell_{\max}^{(\mathrm{sample})}
\leq
D_{\mathcal C},
\end{align}
with equality if and only if the sample contains at least one pair of nodes
whose shortest-path distance equals $D_{\mathcal C}$.

By construction, at least $99.9\%$ of the sampled node pairs have shortest-path lengths not exceeding $C^\star$ in the initial giant component. This coverage applies to the initial sample and does not guarantee the same fraction throughout the SPP dynamics, because edge removal changes shortest-path distances and network connectivity. We therefore evaluate the suitability of this cutoff by comparing the degradation curves obtained with $C=C^\star$ and $C=N$.

\begin{figure}[!htbp]
  \centering
\includegraphics[width=\linewidth]{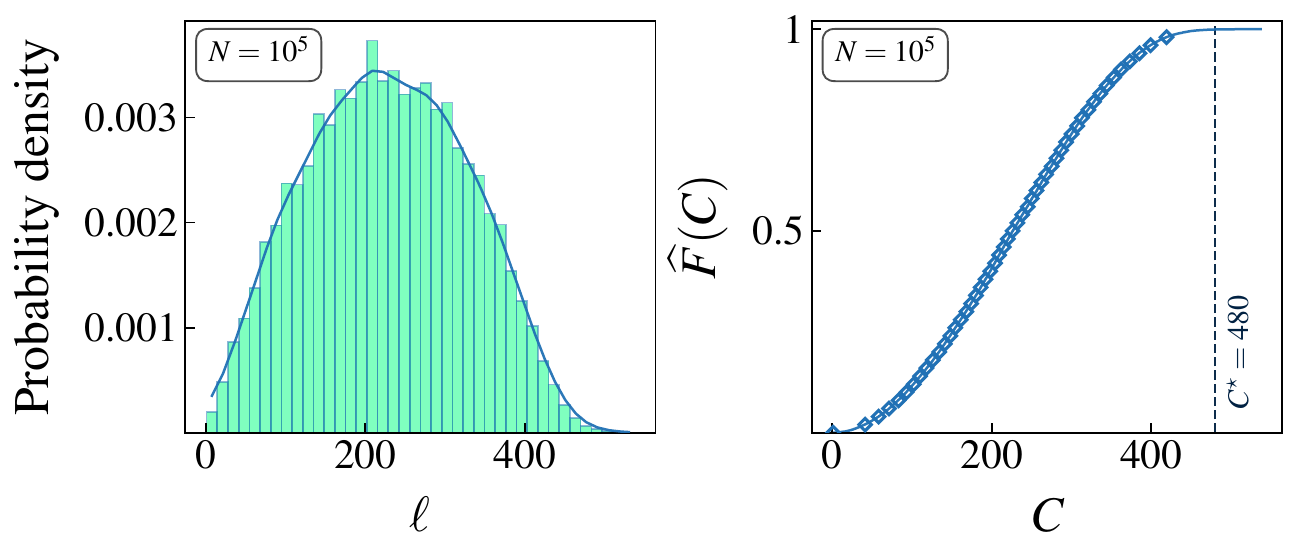}
\caption{Distribution of sampled shortest-path lengths in the giant component of a network with $N=10^{5}$ nodes. (left) Empirical probability density of the sampled shortest-path lengths $\ell$ obtained from uniformly sampled pairs of nodes. (right) Cumulative distribution function $\widehat{F}(C)=\widehat{f}(C)$, representing the estimated fraction of node pairs whose shortest-path length does not exceed $C$. The vertical dashed line indicates $C^{\star}=480$, defined as the smallest integer satisfying $\widehat{f}(C^{\star})\geq0.999$. Consequently, approximately $99.9\%$ of the sampled node pairs satisfy $\ell\leq C^{\star}$, whereas only about $0.1\%$ have shortest-path lengths greater than $C^{\star}$.} 
\label{fig:hist_c_equivalent}
\end{figure}

The procedure described above was applied to estimate the empirical distribution of shortest-path lengths in the giant component. For each network realization, the sample size was chosen as $M=N$, where $N$ is the number of nodes in the network. The resulting empirical distribution and its corresponding empirical cumulative distribution function are shown in Fig.~\ref{fig:hist_c_equivalent}. The equivalent cost $C^{\star}$ is then defined as the smallest positive integer satisfying
\begin{align}
\widehat{f}(C^{\star}) \geq 0.999,
\end{align}
and is subsequently evaluated as a finite-cost approximation to the unrestricted case $C=N$ by comparing the corresponding connectivity degradation curves.

\begin{table}[b!]
\centering
\setlength{\tabcolsep}{16pt}
\renewcommand{\arraystretch}{1.3}
\begin{tabular}{c c}
\hline\hline
$C$ & $\widehat{f}(C)$ \\
\hline
$50$     & $0.02999$ \\
$100$    & $0.11737$ \\
$200$    & $0.41160$ \\
$226$    & $0.50208$ \\
$362$    & $0.90047$ \\
$392$    & $0.95079$ \\
$436$    & $0.99001$ \\
$480$    & $0.99903$ \\
$543$    & $1.00$ \\
$10^{5}$ & $1.00$ \\
\hline\hline
\end{tabular}
\caption{Empirical cumulative distribution $\widehat{f}(C)$ of shortest-path lengths for selected path-length limits $C$ in a quantum internet network with $N=10^{5}$ nodes and initial average degree $\langle k\rangle\approx4$. Each value gives the fraction of $M=N$ unordered node pairs, sampled uniformly without replacement from the initial giant component, whose shortest-path length does not exceed $C$. The equivalent cost $C^\star=480$ is the smallest integer attaining the target coverage $\widehat{f}(C^\star)\geq0.999$.}
\label{tab:cmin_coverage}
\end{table}

\begin{figure}[!htbp]
  \centering
\includegraphics[width=\linewidth]{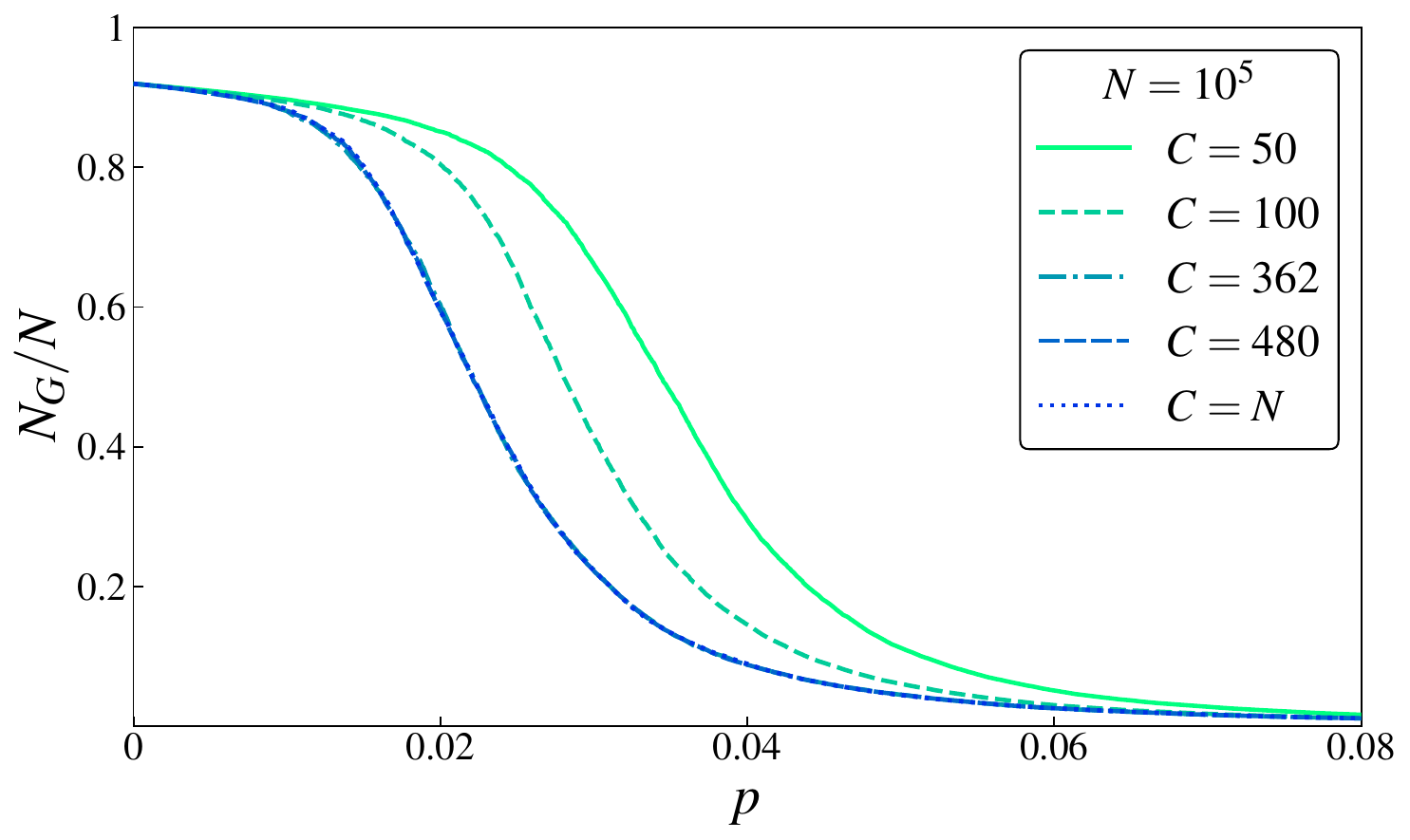}
\caption{Detailed view of the initial connectivity degradation under SPP dynamics with random pair selection for quantum internet networks with $N=10^{5}$ nodes and approximately constant initial average degree $\langle k\rangle\approx4$. The relative size of the giant component, $N_G/N$, is plotted against the fraction of removed edges $p$ for maximum shortest-path lengths $C=50$, $100$, $362$, $480$, and $N$. The equivalent cost $C^\star=480$ reproduces the network degradation observed in the unrestricted case $C=N$, while smaller cutoffs can also yield nearly indistinguishable curves. Each curve is averaged over $10^{3}$ independent network realizations.} 
\label{fig:c-otimo-zoom}
\end{figure}

The quantile criterion does not identify a unique path-length limit for reproducing the unrestricted dynamics. As illustrated in Fig.~\ref{fig:c-otimo-zoom}, the averaged connectivity degradation curves obtained with $C=C^\star$ and some smaller values of $C$ nearly coincide with those obtained for $C=N$ within the region shown. These observations support the use of $C^\star$ as an empirical reference for selecting the path-length limit, while indicating that smaller cutoffs may also be adequate for the macroscopic observable considered.

Here we adopt the criterion $\widehat{f}(C^{\star})\geq0.999$, which corresponds to a highly restrictive coverage level, for which only approximately $0.1\%$ of the pairs of nodes sampled have shortest-path lengths exceeding $C^{\star}$ as shown in Table~\ref{tab:cmin_coverage}. In principle, less restrictive coverage thresholds, such as $0.99$, $0.95$, or $0.90$, can also be adopted, leading to lower values of $C^{\star}$ and, consequently, to a further reduction in computational cost. The choice of the coverage threshold ultimately depends on the desired level of approximation to the unrestricted case.

\begin{table}[b!]
\centering
\setlength{\tabcolsep}{16pt}
\renewcommand{\arraystretch}{1.3}
\begin{tabular}{c c}
\hline\hline
$N$ & $C^{\star}$ \\
\hline
$10^{3}$        & $165$ \\
$2\times10^{4}$ & $227$ \\
$5\times10^{4}$ & $319$ \\
$10^{5}$        & $480$ \\
$5\times10^{5}$ & $993$ \\
$10^{6}$        & $1382$ \\
\hline\hline
\end{tabular}
\caption{Estimated equivalent cost $C^\star$ for quantum internet networks of different sizes $N$ and approximately constant average degree. For each network, $M=N$ unordered node pairs are sampled uniformly without replacement from the initial giant component. The equivalent cost is defined as the smallest integer satisfying $\widehat{f}(C^\star)\geq0.999$, corresponding to the empirical $0.999$-quantile of the sampled shortest-path lengths.}
\label{tab:cmin_by_N}
\end{table}

For each system size $N$, we used a single quantum network realization with initial average degree $\langle k\rangle\approx4$. We sampled $M=N$ unordered node pairs uniformly without replacement from its initial giant component and calculated the empirical $0.999$-quantile of their shortest-path lengths. Table~\ref{tab:cmin_by_N} reports the resulting estimates of $C^\star$. Among the models considered to describe the dependence of $C^\star$ on $N$, we find that the best fit is obtained for a power law with a constant offset:
\begin{equation}
C^\star(N)=a+bN^\beta,
\label{eq:C_star}
\end{equation}
where the fitted parameters are $a=117.32$, $b=0.4618$, and $\beta=0.5736\pm0.0494$. The coefficient of determination is $r^2\approx1$ and the root-mean-square error is $\mathrm{RMSE}\approx22.00$. This estimate is consistent with an approximately square-root dependence over the investigated range of system sizes. When the node density and average degree are kept approximately constant, the radius of the spatial domain grows as $R\sim \sqrt{N}$, while the characteristic length of individual photonic links remains finite. This provides a geometric motivation for the observed growth of $C^\star$, although it does not establish its asymptotic scaling.

These results suggest that the observed growth of $C^\star$ is consistent with an approximately square-root dependence on the system size. That is, the fitted behavior is clearly sublinear, implying that the ratio $C^\star/N$ decreases as the network grows. Consequently, the maximum shortest-path length required to reproduce the unrestricted dynamics represents an increasingly smaller fraction of the theoretical upper bound $N$.
These results support using a finite path-length limit that grows much more slowly than the system size to approximate the macroscopic degradation of the unrestricted process.
By restricting the depth of shortest-path searches, this limit can also shorten simulation time, since the costs $C = C^\star$ and $C = N$ yield approximately the same percolation dynamics.

\section{Emergence of a common crossing point in random networks}\label{sec}

\begin{figure}[!htbp]
\centering
\includegraphics[width=\linewidth]{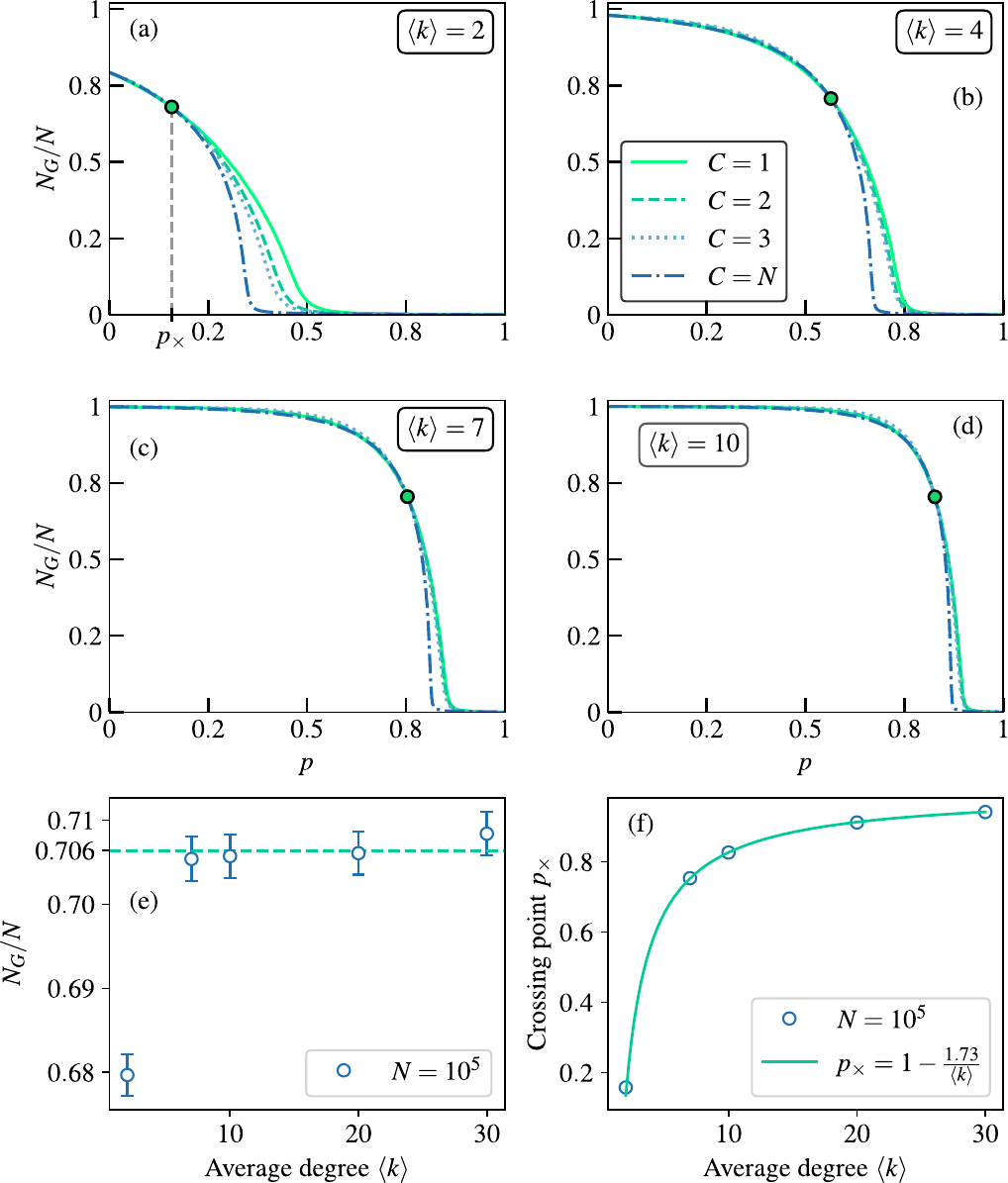}
\caption{Approximately common crossing point in the SPP dynamics on ER networks with $N=10^{5}$. (a–d) Relative size of the giant component, $P=N_G/N$, as a function of the fraction of removed edges $p$, for initial average degrees $\langle k\rangle=2$, $4$, $7$, and $10$, respectively. The curves correspond to $C=1$, $2$, $3$, and $N$ and are averaged over $10^{3}$ independent network realizations. Green markers indicate the narrow regions where the curves for different costs intersect. (e) Giant-component size at the crossing, $P_\times=N_G(p_\times)/N$, as a function of the initial average degree. The results approach an approximately constant value, $P_\times\approx0.706$, indicated by the orange dashed line. (f) Crossing position $p_\times$ as a function of the initial average degree. Symbols represent numerical estimates, and the solid orange curve shows the empirical relation $p_\times\approx1-z_\times/\langle k\rangle$. Here, $z_\times$ is the approximately constant residual average degree inferred from the crossing height through $z_\times=-\ln(1-P_\times)/P_\times$, using the ER giant-component equation for $C=1$ [Eq.~\eqref{eq:er_giant_crossing}].}
\label{fig:fig_crossing}
\end{figure}

Let us now investigate an interesting feature of the shortest-path
percolation dynamics that is already visible in the original results
of Ref.~\cite{Kim_2024}, although it was not explicitly discussed
there. When the relative size of the giant component is plotted as a
function of the fraction of removed edges $p$, the degradation curves
obtained for different values of the cost parameter $C$ intersect
within a narrow region. This defines an approximately common crossing
point, which we denote by $p_\times$ (see Fig.~\ref{fig:fig_crossing}; the phenomenon is also visible in the right panel of Fig.~\ref{fig:spp-quantum-cluster-random-er}). In other words, there exists a
particular stage of the degradation process at which the macroscopic
connectivity of the network becomes nearly insensitive to the cost
scale $C$ that constrains the length of the shortest paths that can be
removed. The existence of such a point is not obvious, since different
values of $C$ generally generate distinct percolation patterns and,
consequently, different network structures at the same fraction of
removed edges.

To characterize this phenomenon and determine how it depends on the
connectivity of the initial network, we consider ER random networks
with fixed size $N=10^5$ and different values of the initial average
degree $\langle k\rangle$. Figure~\ref{fig:fig_crossing} shows the relative
size of the giant component, $N_G/N$, as a function of $p$ for
$\langle k\rangle=2$, $4$, $7$, and $10$, and for the cost values
$C=1$, $2$, $3$, and $N$. For each value of $\langle k\rangle$, the
curves corresponding to different values of $C$ follow distinct
trajectories during the degradation process but approach one another
and intersect within a narrow range of $p$.

The dependence on $C$ is particularly evident for the more sparsely
connected networks, with $\langle k\rangle=2$ and $4$. In these cases,
the unrestricted dynamics $C=N$ begins to reduce the giant component
at smaller values of $p$, whereas the finite-cost cases $C=1$, $2$,
and $3$ remain closer to one another over a larger portion of the
degradation process. As the initial average degree increases to
$\langle k\rangle=7$ and $10$, the differences among the curves become
progressively smaller. The various cost values then produce very
similar degradation curves over most of the range of $p$, although
small differences remain visible in the region where the giant
component decreases rapidly.

Interestingly, despite these differences in the evolution of the
network, the curves for different values of $C$ intersect at
approximately the same value of $p$ for a given initial average degree.
These crossing regions are indicated by the green dots in
Fig.~\ref{fig:fig_crossing}. We obtain approximately
$p_\times\approx0.16$, $0.56$, $0.75$, and $0.82$ for
$\langle k\rangle=2$, $4$, $7$, and $10$, respectively. Thus, the
location of the crossing shifts systematically toward larger fractions
of removed edges as the initial average degree increases.

These results reveal two related effects. First, the influence of the
cost parameter $C$ on the degradation dynamics becomes progressively
weaker as the initial connectivity of the network increases. Second,
and more remarkably, the dependence on $C$ nearly disappears at a
particular stage of the degradation process, giving rise to the common
crossing point $p_\times$. This suggests that $p_\times$ is associated
with a characteristic state of the residual network rather than with
a particular value of the path-length constraint. The systematic
displacement of this point with $\langle k\rangle$ further suggests
that its location is primarily controlled by the connectivity of the
network.

Indeed, a simple empirical relation governing the dependence of $p_\times$ on
$\langle k\rangle$ can be obtained by noticing in the simulations that
the size of the giant component at the crossing point converges to an
approximately constant value as $\langle k\rangle$ increases
[see Fig.~\ref{fig:fig_crossing}(e)]. Denoting this limiting value by $P_\times = N_G(p_\times)/N$, we find
numerically $P_\times \approx 0.706$. Since the curve with $C=1$
corresponds to ordinary bond percolation on an ER network, the size $P$
of the giant component satisfies~\cite{newman2018networks,Newman_2001}
\begin{equation}
P = 1-\exp(-zP),
\label{eq:er_giant_crossing}
\end{equation}
where $z$ is the average degree of the residual network. At the crossing
point, we have
$
z_\times
=
\langle k\rangle(1-p_\times),
$
because a fraction $p_\times$ of the initial edges has been removed.
Evaluating Eq.~\eqref{eq:er_giant_crossing} at $P=P_\times$ gives
$
z_\times
=
-\ln(1-P_\times)/P_\times, 
$
and using $P_\times\approx0.706$, we obtain $z_\times\approx1.73$.
Therefore, if the crossing occurs at an approximately constant
giant-component size, it also occurs at an approximately constant
residual average degree. Writing $p_\times$ in terms of $z_\times$ and using $z_\times\approx1.73$ gives the empirical relation
\begin{equation}
p_\times(\langle k\rangle)
\approx
1-\frac{1.73}{\langle k\rangle}.
\label{eq:px_empirical}
\end{equation}
This relation provides a simple description of the systematic shift of
the crossing point toward larger values of $p$ as the initial average
degree increases. As we see in Fig.~\ref{fig:fig_crossing}(f), numerical results agree well with the empirical law given by Eq.~\eqref{eq:px_empirical}. Further analytical work is needed to explain the origin of this approximately common crossing point and the mechanisms governing its emergence in the SPP dynamics.

\section{Conclusion}
\label{sec:conclusion}

In conclusion, our findings can be compared directly with earlier results on quantum networks. For satellite-based quantum networks, previous work has shown that hubs induce small-world properties and increase robustness against random failures, while attacks directed at these hubs rapidly destroy connectivity~\cite{Brito_2021}. For optical fiber-based quantum communication networks, a phase transition associated with the formation of a giant connected component occurs when the density of nodes becomes larger than a critical value~\cite{Brito_2020}.

We have shown that changes in the spatial structure of the network with system size have a strong influence on the SPP dynamics. When the spatial domain is fixed, larger values of $N$ increase both the node density and the average degree, thereby increasing network robustness. By rescaling the domain, we can instead investigate fragmentation in growing networks with an approximately constant initial average degree. In the context of the quantum internet, this process corresponds to a progressive loss of the routes available for distributing entanglement. The consumption of shortest paths reduces the number of routes through which new entangled pairs can be established between distant nodes, and consequently weakens the capacity of the network to support large-scale quantum communication.

Our comparison with the SPP results of Ref.~\cite{Kim_2024} reveals the same qualitative difference between the gradual loss of connectivity for finite costs and the more abrupt fragmentation observed in the unrestricted case. In the homogeneous random networks considered in Ref.~\cite{Kim_2024}, however, the transition window for unrestricted dynamics becomes narrower as the system size increases, whereas the behavior found here depends on the spatial construction of the network. We do not observe clear convergence of this window when the domain is fixed. For an approximately constant average degree, the window becomes narrower with increasing $N$, in qualitative agreement with the results of Ref.~\cite{Kim_2024} for ER networks.

The authors of Ref.~\cite{Meng_2025} have shown that sparse quantum communication networks are vulnerable to the successive removal of communication paths. In their model, each communication event consumes all entangled links along the selected path, regardless of its length—in contrast to the finite-cost dynamics considered here—leading to a rapid loss of communication capacity in ER networks with finite average degrees, two-dimensional networks, and satellite-based networks~\cite{Meng_2025}. Our results are consistent with this behavior in the unrestricted case, $C=N$, in which any connected pair can communicate along its shortest path and several links can be removed in a single event. For both spatial constructions considered here, this dynamics leads to a faster loss of connectivity than that observed for finite values of $C$. We also find that the response of the network depends on how the average degree changes with system size. When the average degree increases with $N$, the network becomes more robust, although the unrestricted case remains the most destructive.

We have also introduced the equivalent cost, $C^{\star}$, which we estimate by Monte Carlo sampling of the shortest-path length distribution. Within the statistical precision of the simulations, this finite bound on path length leads to the same evolution of the giant component as in the unrestricted dynamics. For networks with an approximately constant average degree, the increase in this cost is approximately consistent with square-root scaling. The finite cutoff $C^\star$ can therefore reduce simulation time by restricting the depth of shortest-path searches, while leaving the percolation dynamics essentially unchanged when compared with the unrestricted dynamics.

In quantum communication networks, entanglement swapping consumes the entangled links used to connect distant nodes~\cite{pant2019routing}. In the SPP model studied here, this consumption is represented by the removal of a shortest path after each successful communication request, leaving fewer routes available for subsequent requests. The bound $C$ sets the maximum number of links that can be consumed in one request. That is, finite values restrict communication to pairs connected by sufficiently short paths, whereas $C=N$ allows any connected pair to communicate along its shortest path. Our results show that this unrestricted use of paths leads to the fastest loss of connectivity, and that the network response depends on the spatial structure and on how the average degree changes with system size. Thus, in the resource-depletion regime considered here, allowing communication over longer paths comes at the cost of consuming more entangled links per successful request.

Finally, we have found that the degradation curves of ER networks for different path-length bounds pass through an approximately common crossing point. The location of this point depends on the initial average degree according to the empirical relation $p_\times\approx1-1.73/\langle k\rangle$, which corresponds to an approximately constant residual average degree at the crossing. Thus, at this stage of the SPP dynamics, the macroscopic connectivity depends only weakly on the imposed path-length bound. An analytical explanation of the crossing, together with a determination of the range over which the empirical relation holds, would be an interesting topic for future research.

\section*{Acknowledgments}

T.P. acknowledges FAPESP (Grant No. 2023/07481-6), CNPq/Brazil (Grant No.~310248/2023-0), and INCT-NeuroComp (CNPq Grant No.~408389/2024-9). This research was carried out using the computational resources of the Center for Mathematical Sciences Applied to Industry (CeMEAI) funded by FAPESP (Grant No. 2013/07375-0). The authors are also grateful to Haroldo V. Ribeiro, Paulino R. Villas-Boas, Diogo O. Soares-Pinto, and Tiago Martinelli for fruitful discussions. 

\section*{Data accessibility}

The Python implementations developed in this work, as well as additional node-pair selection strategies for the SPP dynamics, are publicly available in an open-access repository. The source code ensures full reproducibility of the experiments and provides editable mechanisms for defining the node-pair selection protocol. In particular, the framework supports multiple strategies, including \textit{random}, \textit{hub}, \textit{closeness}, and \textit{domirank} selection~\cite{engsig2024domirank}. The implementation is available at \url{https://github.com/VictorFrancheto/shortest-path-percolation-implementation}.

\bibliography{aipsamp}

\end{document}